\documentclass[]{article}

\usepackage{graphicx}
\usepackage[caption=false]{subfig}
\usepackage{multirow}
\usepackage{booktabs}
\usepackage{hyperref}
\usepackage{array}
\usepackage{capt-of}

\RequirePackage{etex} 
\usepackage{enumitem}

\usepackage[title]{appendix}

\usepackage[backend=biber,
url=false,
doi=false,
eprint=false,
isbn=false,
giveninits,
uniquename=init,
citestyle=authoryear-comp,
uniquename=false,
maxcitenames = 1,
minbibnames = 6,
maxbibnames = 6,
bibstyle=apa]{biblatex}
\DeclareSourcemap{
	\maps[datatype=bibtex]{
		\map{
			\step[fieldset=issn, null]
			\step[fieldset=url, null]
			\step[fieldset=doi, null]
			\step[fieldset=eprint, null]
		}
	}
}

\usepackage{xargs} 
\usepackage[pdftex,dvipsnames]{xcolor}  

\usepackage{xcolor}
\usepackage{tikz-network}
\usepackage{tikz}

\usepackage{algorithm}
\usepackage[noend]{algpseudocode}

\usepackage{amsmath}
\usepackage{amssymb} 

\definecolor{frenchblue}{rgb}{0.0, 0.45, 0.73}
\definecolor{etonblue}{rgb}{0.59, 0.78, 0.64}
\definecolor{darkcoral}{rgb}{0.8, 0.36, 0.27}
\definecolor{urobilin}{rgb}{0.88, 0.68, 0.13}
\definecolor{wisteria}{rgb}{0.79, 0.63, 0.86}

\graphicspath{{plots/}{img/}}

\usepackage[affil-it]{authblk}

\title{Semantic and Relational Spaces in Science of Science: Deep Learning Models for Article Vectorisation}

\author[1]{Diego Kozlowski\thanks{\texttt{diego.kozlowski@uni.lu}; corresponding author}}

\author[2]{Jennifer Dusdal}
\author[1]{Jun Pang}
\author[1]{Andreas Zilian}

\affil[1]{Faculty of Science, Technology and Medicine, University of Luxembourg\\
	L-4364 Esch-sur-Alzette, Luxembourg}

\affil[2]{Faculty of Humanities, Education and Social Sciences, University of Luxembourg\\
	L-4364 Esch-sur-Alzette, Luxembourg}

\date{}

\begin{document}
 	\maketitle
 	\begin{abstract}
 		Over the last century, we observe a steady and exponentially growth of scientific publications globally. The overwhelming amount of available literature makes a holistic analysis of the research within a field and between fields based on manual inspection impossible. Automatic techniques to support the process of literature review are required to find the epistemic and social patterns that are embedded in scientific publications. In computer sciences, new tools have been developed to deal with large volumes of data. In particular, deep learning techniques open the possibility of automated end-to-end models to project observations to a new, low-dimensional space where the most relevant information of each observation is highlighted. Using deep learning to build new representations of scientific publications is a growing but still emerging field of research. The aim of this paper is to discuss the potential and limits of deep learning for gathering insights about scientific research articles. We focus on document-level embeddings based on the semantic and relational aspects of articles, using Natural Language Processing (NLP) and Graph Neural Networks (GNNs). We explore the different outcomes generated by those techniques. Our results show that using NLP we can encode a semantic space of articles, while with GNN we are able to build a relational space where the social practices of a research community are also encoded. 
 	\end{abstract}
 
 	{\bf keywords:} Embeddings, Science of Science, Deep Learning, Graph Neural Networks, Semantic Space, Relational Space
	 
	\section{Introduction}
	\label{sec:intro}
	
	The relation between two research articles is a multidimensional phenomenon. Research articles can be related because of their topics, authors, or the organisational affiliation of the authors. This implies that there is no unique measurement to compare the relatedness of scientific publications. 
	
	When a human compares a pair of articles, he or she can recognise the relatedness in its complexity, given his/her own biases and expertise in the field. The relatedness of a pair of articles is not only interesting for a pairwise comparison, but also to build a holistic representation of a field of research or a discipline. To handle the massively increasing volume and production rate of scientific research articles, we discuss automatised ways to relate research articles to each other. 
	
	One of the most important dimensions of relatedness is the semantic content of articles. Using text as data, Natural Language Processing (hereafter NLP) studies the ways in which textual meaning can be extracted from the documents within a text corpus, in a summarised way~\parencite{jurafsky_speech_2008}. This operationalises the concept of semantic relatedness~\parencite{mikolov_linguistic_2013}. One of those techniques is Topic Modelling, which uses the co-occurrences of words in documents to detect the distribution of topics in a given corpus~\parencite{blei_latent_2003}. \cite{daenekindt_mapping_2020} analysed the field of higher education using correlated topic modelling~\parencite{blei_correlated_2007}, which allowed the authors to compare 17,000 abstracts published in journals in the field of higher education research. \cite{schwemmer_methodological_2020} studied the methodological divergence in sociology between quantitative and qualitative analysis in more than 8,700 research articles in top journals of the field, using the \textit{wordfish} model~\parencite{slapin_scaling_2008}. Another study analysed more than 20 million article titles to investigate the expansion of cognitive boundaries in physics, astronomy, and biomedicine. Findings include  that the number of publications grows exponentially, but that the space of ideas expands linearly~\parencite{milojevic_quantifying_2015}.
	
	Another important aspect in the relatedness of articles is the overarching network structure of science~\parencite{boyack_co-citation_2010}. A network of articles can be made explicit by their references to previous work. Considering direct citations, two articles are linked if one of them contains a reference to the other. This network has a strong temporal dependency, as nodes can only have outgoing links to older articles. The links can also be based on co-citations~\parencite{small_co-citation_1973, kessler_bibliographic_1963}. The network structure has the property of defining the distance between two articles by the length of the path between them, i.e., the number of articles needed to to get from one to another, only following the references lists of the corresponding article at each step. Intertwined with the articles' network, a collaboration network of authors can be created~\parencite{Moody2004}.

	New deep learning models promise to revolutionise the way we approach both, text and network data. Nevertheless, there is an ongoing debate in the Artificial Intelligence community on the bias introduced by algorithms, e.g. \cite{buolamwini_2018, Bolukbasi_2016, whittaker2018ai, Caliskan183}. This implies that, when we introduce new methodologies, and specially black-box models like in deep learning, it is important to study the implicit biases they carry. 
	
	Together with other deep learning techniques, embeddings are an important breakthrough in the field of machine learning. An embedding is a low dimensional dense vector, i.e., of real-valued representations that encodes relevant information from the original, high dimensional data~\parencite{mikolov_efficient_2013}. Nevertheless, not much research has been carried out to bring these techniques closer to the field of Science of Science, except of the following examples: A recent article used Word2Vec~\parencite{mikolov_efficient_2013} to distinguish the most relevant terms in quantitative and qualitative research in the field of Science of Science~\parencite{kang_against_2020}. Paper2Vector, \cite{zhang_p2v_2019}, is a model that trains word-word, document-word, and document-document relations based on the Skip-Gram model~\parencite{mikolov_efficient_2013}. This model, although it uses textual content as well as the citation network, cannot use both at the same time as well as it cannot use other metadata features. Another approach used Graph Neural Networks (hereafter GNN) models, a deep learning model that leverage on the network structure of the data, as well as the BERT pre-trained model for the textual embedding, which constitutes the state of the art in NLP, but it does not consider the possibility of using the textual embedding as input for the GNN~\parencite{jeong_context-aware_2020}. 
	
	We have selected Science of Science as a case study to explore these new methodological approaches because it is a highly complex and multidisciplinary field of research~\parencite{Fortunato2018} that aims to explore the driving forces of science and to develop new methods to better understand its evolution over time. The emergence of the field has speed up with the availability of large-scale data sets on science production and the disruption of disciplinary boundaries, that encouraged scientists from different disciplines to closely collaborate, an ideal playground to study different types of embeddings as an innovative tool for the methodological development in Science of Science, for an in-depth analysis scientific research articles. We identified this as an important research gap in this developing field of research, and will discuss potential uses and limitations of these new methods. The article focuses on document-level embeddings based on the semantic and relational aspects of articles, using NLP and GNN's to explore semantic and relational spaces in Science of Science. Four different aspects will be analysed: (a) collaboration-patterns~\parencite{Sooryamoorthy2009}; (b) the cumulative effect on citations, i.e., the Matthew effect~\parencite{de1963little,garfield1972citation, garfield1979citation}; (c) the position of countries in science global production~\parencite{King2004}; and (d) the quantitative-qualitative divide~\parencite{kang_against_2020}.
	
	Our main hypothesis is that, while textual embeddings build a representation of the semantic space, GNN embeddings focus on the relational and structural space of a network of research articles. Therefore, textual embeddings help to identify similar content, whereas GNN embeddings are useful to study the embedded social relations in the production of scientific knowledge. An in-depth investigation of this hypothesis is an important step for Science of Science as a field of research, allowing us to pair powerful analysis techniques from computer science  with a thematic disciplinary foundation.
	
	The main objective of this methodological contribution is to present a approximation to the use of embeddings in the field of Science of Science. We propose two families of models that use different types of inputs and generate different types of insights: First, we build articles embeddings based on their textual characteristics, including titles, keywords, and abstracts. For this family of models, we use three different techniques: Topic Modelling~\parencite{blei_latent_2003}, Doc2Vec~\parencite{mikolov_distributed_2013}, and BERT~\parencite{devlin_bert_2019}. Doc2Vec was selected because it is specifically designed for document-level embeddings. BERT achieves the state of the art for various NLP tasks~\parencite{tenney-etal-2019-bert}. For a non-deep learning benchmark, we selected the Topic Modelling approach, which is a widely used framework. Second, we build GNN models that include text, metadata and the citation network of selected articles. The GNN models are trained on the link prediction task, i.e., predicting whether two articles are linked by a citation, and therefore focus on the network properties. This study does not develop new methodologies from text or network embeddings, but intends to act as a bridge between the new developments made on the field of Deep Learning, and studies on Science of Science. In our case study, we try the different models on a data set of 22,151 articles from Science of Science, involving different fields, ranging from history and philosophy of science to library and information sciences.

    The following two research questions are guiding our analysis: How can we encode the \textit{relational} dimension of articles, and which are its properties? How can we encode the \textit{semantic} dimension of articles, and which are its properties? 
     
	\smallskip
	\noindent{\bf Structure of the paper.}
	First, we will present the data set characteristics, while in Section~\ref{sec:Embeddings} we provide an overview of embedding techniques in both text and networks, the experimental setup and the performance metrics that are used to evaluate the models. In Section~\ref{sec:Results} we present our results. We finish the paper with a conclusion and final remarks for future research in Section~\ref{sec:Conclusions}.
	
	\section{Data Set and Network Statistics}
	\label{sec:Data}
	The data set was built on a ``journal-based'' approach. First, we defined a set of core journals in the field of Science of Science. This selection is based on a recommendation of journals of the ``International Society for Scientometrics and Informetrics'' (ISSI)\footnote{\url{http://www.issi-society.org/links/}, last accessed October 6th, 2020}, and has been expanded to include related journals from social sciences, and history and philosophy of science to show the wide variability and multidisciplinarity of this field of research. Our main goal was to include all journals that focus exclusively on Science of Science, independently of their disciplinary approach. Second, from this selected set of journals, we included all articles that are available in Elseviers' Scopus journal database.  Methodologically, we have used Scopus API\footnote{Data collection between April 8 and July 30, 2020, via \url{http://api.elsevier.com}} to extract the data. By including all articles from these journals, we avoided a potential selection bias towards keywords, and ensure a comprehensive investigation of the field of Science of Science. Given that this study focuses on the methodological aspects of the use of embeddings, we decided to work with this non-exhaustive corpus of research articles, which gives us the advantage to investigate the distribution of embeddings at the journal level. As in Section~\ref{sec:Differences} we carry out a country level analysis, it is important to mention the limitations of the Scopus database. Limitations of the data set include a bias towards English-speaking and Western-oriented journals, and a lack of coverage of social sciences. We are only including peer-reviewed articles and do not investigate other publication formats (e.g., monographs, contributions to edited volumes, conference proceedings, etc.).
	
	Table~\ref{table:data_stats} displays information on the journals in our sample, including the number of articles retrieved, the mean and maximum number of citations by journal, and the year of the first and last publication. We have used ``Scopus Subject Areas and All Science Journal Classification Codes'' (ASJC) for a first differentiation of the journals by discipline\footnote{\url{https://service.elsevier.com/app/answers/detail/a_id/15181/}}. Having investigated the repetition of these areas in the different journals, and additionally based on the results of Topic Modelling (see Section~\ref{sec:LDA}), we assigned each of them to four fields of study: \textit{Management}, \textit{Library and Information Sciences}, \textit{History and Philosophy of Science} and \textit{Other Social Sciences}, composed by \textit{Education, Communication and Anthropology}. We consider the repetition of areas and acknowledge that some journals are more multidisciplinary in nature than others, which could lead to an assignment to another field. For example, \textit{Social Studies of Science} has as subject areas \textit{History}, \textit{Social Sciences (all)} and \textit{History and Philosophy of Science}, so it could be assigned to both, \textit{Other Social Sciences} and \textit{History and Philosophy of Science}. This implies that the defined fields cannot be perfectly matched and characterise each journal. In addition, fields are not equidistant from each other. For example, the relation between \textit{Management} and \textit{Library and Information Sciences} is closer than the relationship between those and \textit{Philosophy}. Methodologically, these fields are not used as features in the subsequent models, so they do not introduce biases to the models, but are a helpful tool used for the analysis of results, as they allow to visually study the projection of the embeddings in Sections~\ref{sec:relational_space}~and~\ref{sec:semantic_space}, and for studying the journals epistemic practice division in Section~\ref{sec:quant_qual_divide}. In Section~\ref{sec:LDA}, we show that we can partially infer these fields from the Topic Modelling results.
	
	The distribution of the number of research articles and citations per journal is skewed (see also \cite{bornmann2008citation}), with a tendency towards \textit{History and Philosophy of Science} having less citations per article than the rest of the fields, which corresponds to different citation and publication behaviours across disciplines~\parencite{lillquist2010discipline}. Overall, the data set includes 22,151 articles, with an average of 20.7 citations per articles.
	
	\begin{table}[!t]
		\caption{Statistics of the data set.}
		\label{table:data_stats}

		\resizebox{\textwidth}{!}{%
			\begin{tabular}{ll@{\hspace{-.5cm}}rrrrr}
				\toprule
				Field & Journal & \begin{tabular}[c]{@{}r@{}}Articles\\ Retrieved\end{tabular} & \begin{tabular}[c]{@{}r@{}}Mean\\ Citations\end{tabular} & \begin{tabular}[c]{@{}r@{}}Max\\ Citations\end{tabular} & \begin{tabular}[c]{@{}r@{}}First\\ Year\end{tabular} & \begin{tabular}[c]{@{}r@{}}Last\\ Year\end{tabular} \\ \midrule
				\multirow{2}{*}{Management} 
				& Research Policy & 3,221 & 83.75 & 4,820 & 1971 & 2020 \\
				& Science and Public Policy & 1,707 & 13.27 & 462 & 1976 & 2019 \\ \hline
				\multirow{2}{*}{\begin{tabular}[c]{@{}l@{}}Library and\\ Information Sciences\end{tabular}} 
				& Scientometrics & 5,136 & 20.04 & 1,334 & 1978 & 2020 \\
				& Journal of Informetrics & 876 & 22.63 & 352  & 2007 & 2020\\ \hline
				\multirow{8}{*}{\begin{tabular}[c]{@{}l@{}}History and\\Philosophy\end{tabular} } 
				& Synthese & 4,151 & 8.53 & 910 & 1946 & 2020 \\
				& Social Studies of Science & 1,069 & 40.95 & 4,709 & 1971 & 2020\\
				& Science and Education & 1,034 & 11.60 & 298 & 1992 & 2020 \\
				& \begin{tabular}[c]{@{}l@{}}Studies in History and\\ Philosophy of Science\end{tabular} & 911 & 8.76 & 145 & 1974 & 2020\\
				& Isis & 523 & 12.47 & 415 & 1977 & 2020 \\
				& Science, Technology and Society & 345 & 6.07 & 122 & 1996 & 2020\\
				& \begin{tabular}[c]{@{}l@{}}British Journal For\\ the History of Science\end{tabular} & 276 & 9.57 & 88 & 1962 & 2020 \\
				& Science and Technology Studies & 111 & 5.29 & 39 & 2012 & 2019\\ \hline
				\multirow{4}{*}{\begin{tabular}[c]{@{}l@{}}Other Social Sciences:\\Education,\\Communication\\ and Anthropology\end{tabular}} 
				& Public Understanding of Science & 977 & 25.91 & 518 & 1996 & 2020\\
				& Science, Technology and Human Values & 757 & 32.87 & 828 & 1982 & 2020\\
				& Research Evaluation & 666 & 13.15 & 223 & 1991 & 2019\\
				& Minerva & 391 & 16.51 & 624 & 1965 & 2020 \\ \hline
				& Total & 22,151 & 20.71 & 4,820 & 1946 & 2020\\ \bottomrule
			\end{tabular}
		}
	\end{table}
	
\begin{figure}[!t]
		\begin{minipage}[b]{.47\linewidth}
			\begin{tabular}[b]{ll}
				\toprule
				Metric                                           & Value \\ \midrule
				Number of nodes                                 & 16,578 \\
				Number of links                                      & 68,797 \\
				\begin{tabular}[c]{@{}l@{}}Number of nodes in the\\giant component\end{tabular} & 15,615 \\
				\begin{tabular}[c]{@{}l@{}}Number of links in the\\giant component\end{tabular} & 68,168 \\
				Diameter                                         & 37    \\
				Average degree                                   & 8.3   \\
				Max degree                                       & 282   \\
				Cluster Coefficient ($C$)                         & 0.081 \\
				Mean path length ($L$)                            & 6.14  \\
				\begin{tabular}[c]{@{}l@{}}Erdös-Renyi average \\cluster coefficient ($C_r$ )\end{tabular}  &   0.0005    \\
				\begin{tabular}[c]{@{}l@{}}Erdös-Renyi average \\mean path length ($L_r$)\end{tabular}      &     4.83  \\
				$C/C_r$                                           &    162.45   \\
				$L/L_r$                                           &     1.27  \\ \bottomrule
			\end{tabular}
		\end{minipage}
		\begin{minipage}[b]{.51\linewidth}  
			\includegraphics[width=\linewidth]{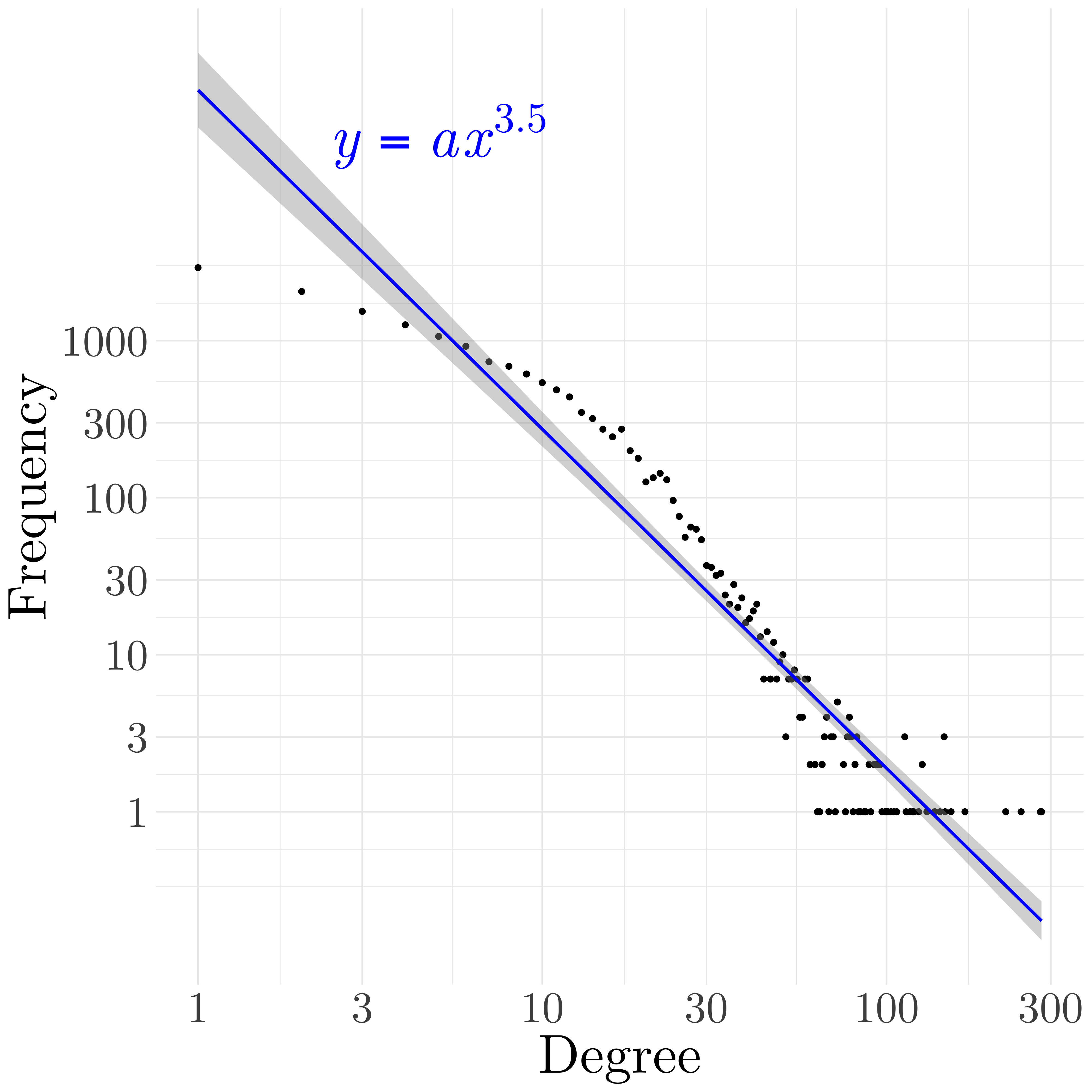}
		\end{minipage}
		\centering
		\captionsetup{justification=centering,margin=2cm}
		\caption{Network Statistics, log-log degree distribution and power law fit.}
		\label{table:network_stats}
	\end{figure}

	From the collection of 22,151 articles, we retrieved the references to build the citation network. 75\% of them either cited or were cited by another article in our sample. This subset builds the basis for the citation network\footnote{In Appendix~\ref{sec:oon_stats}, we show the summary statistics of the articles inside and outside of the network.}. Figure~\ref{table:network_stats} presents a summary of the characteristics of the resulting network the log-log degree distribution of nodes and the fitted power law distribution. The network has 16,578 nodes (articles) and 68,797 links (citations). The average degree of the network is 8.3 compared to 20.7 connections when using the entire Scopus database. While the network is not connected, the giant component includes most of the edges and vertices. We built 100 replications of a random Erdös-Renyi network~\parencite{erdos_evolution_1960} with the same number of vertices and edges, and computed the average cluster coefficient as well as the average mean path length for comparison. We show that the ratio of the network cluster coefficient with respect to the random Erdös-Renyi network is 162, while the ratio between our network mean path length and that of the random Erdös-Renyi network is 1.27. This means that our network has the properties of \textit{small word} networks~\parencite{davis_small_2003,iyer_managing_2006}. 
	
    \section{Methods}
	\label{sec:Embeddings}
	This section introduces different approaches for dense vector representations of documents, called embeddings, which will be used for analyses of scientific research articles in Science of Science. Then, the implementation of details and performance metrics will be presented.

	\smallskip\noindent{\bf Classic Machine Learning Approaches}
	First, we will focus on classic machine learning approaches for encoding research articles	based on feature engineering, to highlight its differences between those and deep learning based approaches. Feature engineering refers to the way in which measurable attributes of an observation can be encoded. In a traditional data set, each observation is defined as a vector, and the full data set is therefore a composition of these vectors, building a matrix of observations as rows and columns as features~\parencite{Broman_2018}. If the data consists of a network, besides the features-matrix, another matrix to describe the network structure has to be implemented, the so-called  \textit{adjacency matrix} $A$~\parencite{barabasi_network_2016}. 
	An article's vector representation can be any metric way to summarise its measurable characteristics. Such a vector can describe its metadata features, for example, the year of publication, the number of citations at a moment in time, the number of authors, or a label assigned to the organisation or journal. It can also include descriptors of the network in which the article is embedded: degree, betweenness, or other centrality measures.
	
	The classic treatment of the textual content of an article, for example, the title, abstract, keywords, or full text, is based on the \textit{bag of words}: A Document-Term Matrix (DTM) where each document is represented as a vector, indicating which of the words in the vocabulary is present in a given document~\parencite{jurafsky_speech_2008}. 
	In the DTM matrix, the $i$-th row represents the $i$-th  document and the $j$-th column represents the $j$-th word in the vocabulary of the corpus, the value $x_{i,j}$ can either indicate if the $j$-th word is present in the $i$-th document as a  binary value, the number of times it appears, or a normalised counting, like Term Frequency-Inverse Document Frequency (TF-IDF)~\parencite{jurafsky_speech_2008}.
	
	As any $n$-dimensional vector, an article constitutes a point in an $n-dimensional$ space. Each of these $n$ dimensions keeps its independent meaning, which is useful for making interpretations over the position of different documents in the space. The representation might be affected by problems of high dimensionality and sparsity. This is specially true for encoding text, as the vocabulary size can rise to tens of thousands of words, with a high probability that most words will not appear in most documents. For research articles, this might be true for features like authors, organisational affiliations, or journals, if each value is encoded as a dummy variable, i.e., as many variables that take only the value 0 or 1 to indicate the absence or presence of each category. In highly dimensional spaces, the notions of distance become blurred and is hard to generate new insights from the data~\parencite{bellman_dynamic_1966}. Therefore, when studying the relations between articles, the analysis tends to focus on a restricted subset of the multiplicity of dimensions that exist. Compared to other methodological approaches, deep learning models can take a multiplicity of dimensions of analysis into account. Applying deep learning methods contributes to development on this research gap, as they are able to use multiple features and select those that are more relevant based on the optimisation problem. 
	
	Once encoded, descriptive statistics can be used to investigate the relations between observations and features. When the number of dimensions to be considered increases, descriptive statistics can not deal with the studied phenomenon, and some types of models might be necessary to assess the importance of different features. Modelling relations is an entrenched way to reduce the complexity of the problem by selecting the dimensions in focus. One possibility is to use models that measure the relation between two aspects of the phenomena, for example, linear models. These types of models demand a lot of time and expert knowledge, as they require a careful design and hand-engineered features to have a better performance. New developments in machine learning, like deep learning models, can learn from the data to encode the most relevant information, allowing researchers to focus on the analysis of the results.
	
	\smallskip\noindent{\bf Deep learning models}
	Deep learning models are designed in an end-to-end way. Their goal is to minimise the feature engineering steps, and to let the model itself define which the most important latent features are. These models have greater flexibility in terms of the inputs they can receive and the outputs they generate~\parencite{Goodfellow-et-al-2016}. In this article, we focus on those models where the output is a subspace projection where observations are represented. This encoding is defined by the models to maximise an associated task that changes between models. If the model works properly, it will encode those aspects from the original data that are most relevant for solving that associated task. Using this new, data-driven way of encoding information, it is possible to generate new insights comparing aggregated levels of representation like countries or journals. For this type of modelling, we consider two sub-types: First, the textual embeddings that use textual features as their input. From this, we will infer the \textit{semantic space} of articles. We aim to encode the conceptual sense of each article as a low dimensional vector. Second, we will train models that use as input both text and the citation networks, from which we will infer the \textit{relational space} of research articles. In the relational space, we aim to encode the citation practices as a social phenomenon into a low dimensional vector. Here we consider the \textit{semantic and relational spaces} as the mathematical objects that represent those properties of articles. The embeddings are the deep learning implementations that we train over our data set to approximate those mathematical objects.
	
	\subsection{The Semantic Space of a Research Article}
	In this section, we explain the three models that we have used to build the semantic space of research articles. Doc2Vec~\parencite{mikolov_distributed_2013} and BERT~\parencite{devlin_bert_2019} are deep learning models based on the Word2Vec model~\parencite{mikolov_efficient_2013}. The Latent Dirichlet Allocation model (hereafter LDA), which is a non deep learning approach, is extensively used for topic modelling, but that can also be considered as an embedding.
	
	\smallskip\noindent{\bf Word embedding.}
	The representation of documents for training deep learning models is commonly based on word embeddings~\parencite{mikolov_efficient_2013, bojanowski_enriching_2017,pennington_glove_2014}. In word embeddings, each word is represented as a dense vector, and documents are a concatenation of those vectors. To build this representation, Mikolov proposes Word2Vec,~\parencite{mikolov_efficient_2013}, and the \textit{Skip-Gram} implementation. Given a corpus of text and a window size, the model defines the context of a word as the surrounding words within the window size. Then, for each word, it tries to predict its context, internally building a vector for each word. When trained, it learns to project closer words with similar meaning. This means that when we use this model on our Science of Science data set, for example, words like \textit{technology} and \textit{innovation} have a closer representation between them than with the word \textit{student}. 
	When using word embeddings, the document is represented as a matrix of word vectors. Our goal is to make an embedding of articles, instead of words. \cite{mikolov_distributed_2013} includes the identifier of the corresponding document as an additional token to the context window. In this way, it creates both an embedding for words and for documents.
	
	\smallskip\noindent{\bf BERT embedding.}
	BERT~\parencite{devlin_bert_2019} improves the Word2Vec model using \textit{attention} mechanisms~\parencite{vaswani_attention_2017}. This means that not every word in the context is equally considered while trying to predict the next word in the context, which implies that the word embedding of a word is also determined by the context. BERT is also based on the principle of transfer learning. This means that the word vectors can be learned on a big general domain corpus, instead of the specific corpus for each task. This is useful because to build a robust representation of concepts, the word embeddings need billions of observations~\parencite{mikolov_efficient_2013}. 
	
	\smallskip\noindent{\bf LDA embedding.}
	For comparison, we also use the LDA model proposed by~\cite{blei_latent_2003}. This model is not based on a deep learning architecture. Instead, it is a generative Bayesian model. It starts from the premise that a corpus is a collection of topics, and that each document is a mixture of those topics. The model takes the distribution of words within documents as input and generates the topics as a probability distribution of words. For example, if in our corpus \textit{science policy} is a recurrent discussion, LDA would define a topic were words like ``technology", ``innovation" or ``policy" are the most relevant. Given these topics as distributions over words, the model will also define each document as a distribution over those topics. LDA is not usually seen as generating an embedding, as this terminology is usually found within the deep learning community. Nonetheless, its output can be thought of as an embedding for articles, as we can use the articles' distribution over topics as their low-dimensional representation\footnote{In Appendix~\ref{sec:lda_def} we give further details of the LDA model.}.
	
	These three models are compared based on their T-SNE projection~\parencite{van_der_maaten_visualizing_2008}, and also based on how much the GNN model improves when using each of these as input. 
	
	\subsection{The Relational Space of a Research Article}
	The semantic embeddings are built to study exclusively the textual content of documents. An analysis of research articles can consider more than its textual content. We have the opportunity to assess other meta-data features and the citation network to make relationships between articles explicit. GNN's have the potential to assess a more holistic representation of research articles.
	
	GNN is a developing field in the deep learning community that tries to apply techniques that have been proven as useful in computer vision and NLP, to problems where the data has a network structure. Multi-Layer perceptrons work well with flat inputs, where it learns the compositionality of features, but it does not consider their specific dependencies explicitly~\parencite{Goodfellow-et-al-2016}. Recurrent Neural Networks (RNN) are designed for sequential inputs, like text, where the order of the input features is explicitly fed to the network~\parencite{sutskever2011generating}. Convolutional Neural Networks (CNN) are useful when dealing with images, where spatial relations within the image are searched, independently of the specific position in the grid~\parencite{LeCun1989}. The problem with graphs is that while they have explicit relations among nodes, which we would like to incorporate to our models, they are not as regular as relations in images or text, so traditional RNN and CNN cannot deal with this type of data structure. The two main differences are:
	
	\begin{enumerate}
		\item nodes have a variable number of neighbours; and 
		\item neighbours do not have a predefined order.
	\end{enumerate}
	
\begin{figure}[!t]
		\centering
		\includegraphics[width=.75\linewidth]{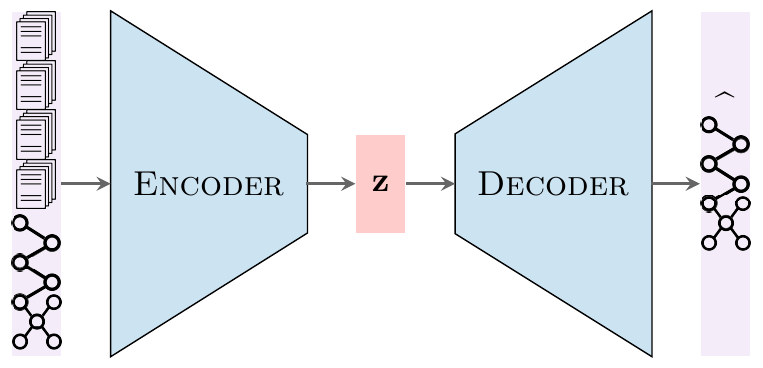}
		\caption{Graph Encoder-Decoder architecture. Author's representation, based on~\cite{Allingham_2020}.}
		\label{fig:gae}
	\end{figure}

	Deep learning on graphs deals with these issues, trying to generalise RNN and CNN to the complex relations in networks. Although GNN can be used for a number of tasks, like node and graph classification, we approach the embedding generation as an unsupervised problem, aiming to reconstruct the node's neighbourhood. This means that we are not trying to predict some node's features while we train our deep learning models, but we try to rebuild the network structure.
	Following~\cite{hamilton_representation_2017}, we approach the problem with an Encoder-Decoder framework.
	Figure~\ref{fig:gae} shows the outline of this architecture.

	From the Encoder-Decoder perspective, the model follows two main procedures: The encoder, $ENC$ takes the nodes features and the network structure and generates a low-level representation of nodes. The decoder, $DEC$ takes the low-level representation as an input and tries to rebuild the network structure.
	Following~\cite{kipf_variational_2016} proposal of the Graph Autoencoder (GAE), the main element of the decoder function is the \textit{inner product} between nodes' low-level representations: 
	$$
	\text{DEC} = ZZ^T, 
	$$
	where $Z$ is the low-level matrix representation of nodes with dimension $n \times d$ with $n$ the number of articles and $d$ the dimension of the article embedding, generated by the encoder. This generates a pairwise decoder, where for each node, we reconstruct its relation with all other nodes, generating an $n \times n$ matrix. If nodes share a similar low-level representation, then the inner product will give a higher value for their pairwise relation.
	
	If on top of the inner product function we apply a sigmoid activation layer, the decoder will produce the pairwise relation of nodes expressed as a probability, i.e., the probability of the two nodes being linked in the network:
	$$
	\hat{A} = \sigma(ZZ^T),
	$$
	where $\hat{A}$ is the reconstructed adjacency matrix.
	The goal is to optimise the encoder in order to minimise the reconstruction loss: 
	$$
	\mathcal{L} = \sum_{(u,v) \in \mathcal{V}}\ell({\hat A}[u,v],A[u,v]),
	$$
	where $\ell$ is the loss function, in our case the binary cross entropy, between the reconstructed link of nodes $u$ and $v$, and the true value in the adjacency matrix. Optimising this loss function implies training an encoder that will generate an embedding representation of nodes that preserve their similarities in terms of the network structure.
	
	The main difference between the used models is how they define the encoder step. We use as encoders the Graph Convolutional Network (GCN)~\parencite{kipf_semi-supervised_2017}, GraphSAGE~\parencite{hamilton_inductive_2017}, Graph Isomorphic Network (GIN)~\parencite{xu_how_2019}, Graph Attention Network (GAT)~\parencite{velickovic_graph_2018}, Attention-based Graph Neural Networks (AGNN)~\parencite{thekumparampil_attention-based_2018} and GraphUNet~\parencite{gao_graph_2019} layers, which constitute the current state of the art in the field. In Appendix~\ref{sec:gnn_models}, we present an overview of these models.
	
	\subsection{Implementation Steps} \label{sec:Setup}

	In this section we present the pre-processing, data cleaning, features, hyperparameters, and network architectures that we have used to build and evaluate the models. 
	
	All textual models were built using a combination of the title, abstract, and keywords suggested by the authors of each article. Given that the words in the title and keywords are good representations of the content of an article, we use them in triplicate: we build a text that has three times the title, three times each keyword and once the abstract. To clean the data, we remove stopwords and trademarks of the journals. After this, we replace the numbers with the special token ``num". We also did stemming, and later replaced the stem with the most frequent word with that stem. 
	
	The Doc2Vec model was implemented using Gensim~\parencite{rehurek_software_2010} with the `distributed memory' learning algorithm, a vector size of twenty and a window size of ten, and using the concatenation of context vectors. To train the BERT embedding, we used the HuggingFace implementation~\parencite{wolf_huggingfaces_2020} with the `bert-base-uncased' pre-trained model. Given that BERT generates word vectors, the sentence embedding was built as the mean of the token embeddings in each sentence. The LDA model was implemented using Scikit-learn~\parencite{pedregosa_scikit-learn_2011} with twenty components, removing tokens that appear in less than five documents or more than 65\% of the documents.
	
	The GNN models were trained using the following features: 
	\begin{itemize}
		\item First author affiliation ID
		\item First author ID
		\item Year of publication
		\item Journal subject area (three per journal, as dummy variables)
		\item Topic distribution (from LDA)
		\item Keywords, title, and abstract  information, either as TF-IDF, or using the sentence embedding from Word2Vec or BERT
		\item Cumulative citations after $t$ years from its publication, for $t \in{1,...,10}$, and total number of citations\footnote{When the article has been published less than  $t$ years ago, the number of citations has been imputed, based on the previous year number of citations, and the mean variation from $t-1$ to $t$.}.
	\end{itemize}
	
	The GNN architectures were implemented using the Pytorch Geometric implementation~\parencite{fey_fast_2019}. In all cases, we use the Graph Autoencoder~\parencite{kipf_variational_2016} with the inner-product decoder. The main difference between the GNN models (GCN, GraphSage, GIN, GAT, AGNN, GraphUNet) is on the encoder side. To find the best set of hyper-parameters, we replicated the specifications from the original papers. 
	The following hyper-parameters have been used in each model: 
	The GCN has an output dimension of 32 and has been built with two GCN convolutions, with a Relu activation layer~\parencite{Agarap2018}. 
	GraphSage was trained with a similar architecture, using the Sage convolutional layer.
	The GIN model was trained with five GIN convolutional layers, using ELU activation layer~\parencite{clevert_fast_2016}, batch normalisation and a normalisation layer in the end. 
	The GAT model was trained with two GAT convolutional layers, with a dropout of 0.6 and a normalisation layer in the end, with an ELU activation layer after the first convolution; the embedding dimension was 16.
	The AGNN model also has two convolutions, starting with a linear projection followed by a Relu activation layer.
	The GraphUNet has the same dropout as the one mentioned in the original paper~\parencite{gao_graph_2019}, with a depth of 4 and an embedding dimension of 16. 
	
	\subsection{Performance Metrics}
	We decided to use two traditionally used metrics in GNN for the evaluation of the link prediction task: the Average Precision (AP) and the Area Under the Receiver Operating Characteristic Curve (AUC). To explain AP, we have to introduce some preliminary notions: for a binary classification model where an observation can be either positive or negative. A True Positive ($tp$) is an observation predicted as positive that is a real positive case. In the same way, we can define true negative ($tn$), false positive ($fp$), and false negative ($fn$). Furthermore, we define 
	$$
	\text{Precision} = \frac{tp}{tp + fp},
	$$
	$$
	\text{Recall} = \text{TPR} = \frac{tp}{tp + fn},
	$$
	$$
	\text{FPR} = \frac{tp}{fp + tn} \,.
	$$
	
	Precision measures describe how many of the predicted links are actual links. Recall, or True Positive Rate (TPR) measure, how many of the actual links are predicted by the model as positive cases. The False Positive Rate (FPR) measures the ``false alarm", i.e., the ratio between true positive and true negative cases. Given that the models predict a ranked sequence of link candidates, i.e., each potential link is associated with a probability. There is an implicit trade-off between Precision and Recall, or between TPR and FPR. For each mode, using the ranked predictions, we can build a curve of Precision against Recall. The AP is the area under that curve, and can be computed as
	$$
	\text{AP} = \sum_n (R_n - R_{n-1}) P_n \,.
	$$
	This means, for every candidate it computes the precision at that point, weighted by the change in the Recall. In a similar way, the AUC is the area under the TPR against FPR curve.
	
	After having presented our data and methods as well as the experimental setup of our study, we now present our results.
	
	\section{Results}
	\label{sec:Results}

	\subsection{Topic Modelling}
	\label{sec:LDA}
\begin{figure}[!t]
		\centering
		\includegraphics[width=\linewidth]{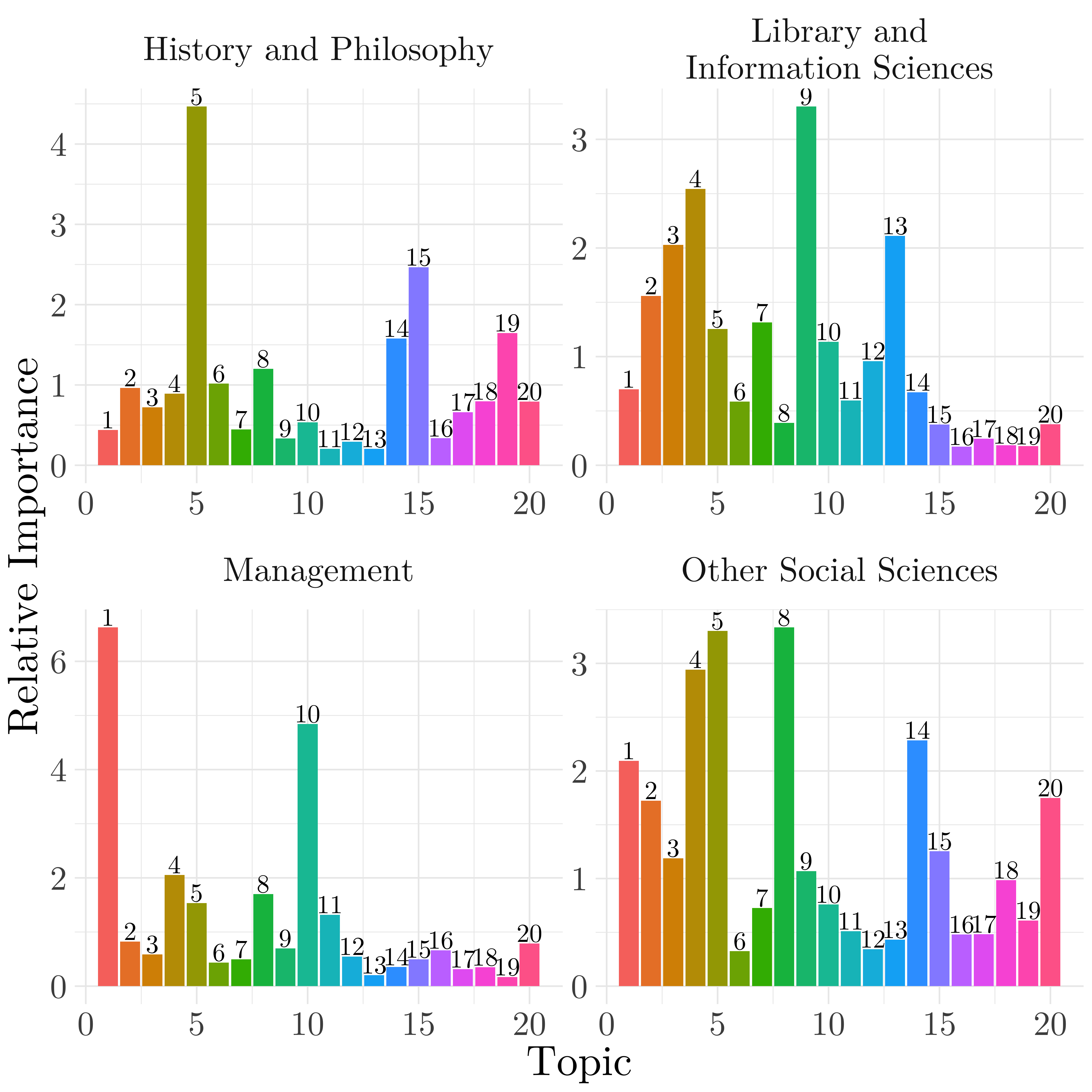}
		\caption{Relative importance of topics per field.}
		\label{fig:topic_modeling}
	\end{figure}

	To present a first characterisation of the data set, we use Topic Modelling to find the latent space of sub-fields within the corpus, using the LDA model~\parencite{blei_latent_2003}. LDA provides two different outputs: First, the distribution of word over topics, which is useful for defining the meaning of each topic. Second, the distribution of topics over documents\footnote{With the distribution of topics over documents, we will later take LDA as a way of embedding over the original space.}. Using LDAvis, we built an interactive visualisation of the distribution of words over topics\footnote{Available at \url{https://diegokoz.github.io/scisci/}}~\parencite{sievert_ldavis_2014}. Figure~\ref{fig:topic_modeling} show the relative importance of each topic per field, calculated as the proportion of the topic in the field over the proportion of the topic in the entire data set. 
	
	
	If we compare the the most relevant words per topic\footnote{See Table~\ref{table:lda_top} in the Appendix} with the distribution per field in Figure~\ref{fig:topic_modeling} we can see that in \textit{History and Philosophy of Science} the topics discussed are \textit{education}~(topic 5), shared also with the field of \textit{Other Social Sciences}, \textit{history}~(topic 15) and \textit{logic}~(topic 19). In the field of \textit{Library and Information Sciences} many different topics are discussed, with especial emphasis in \textit{bibliometrics}~(topic 9), \textit{universities and scientists}~(topic 4), also shared with \textit{Other Social Sciences}. In \textit{Management} we can see the focus in \textit{technology policy}~(topic 1) and \textit{patents and firms}~(topic 10). In \textit{Other Social Sciences} there is also a variety of topics, and besides the shared topics 4 and 5, there is also attention on \textit{public decision making}~(topic 8).
	
	The LDA results show the wide variety of topics covered in the field of Science of Science and also within the different disciplines involved in it. Some topics are shared across fields, and some journals, like \textit{Scientometrics}\footnote{See Figure~\ref{fig:topic_modeling_journals} in the Appendix} cover a wide many thematic studies. This analysis also shows that there is no unique unequivocal way of gathering journals by field. 

	\subsection{Evaluation of Embeddings}
	In this section, we perform an evaluation of the multiple models proposed. For this, we quantitatively compare the performance of the models on the link prediction task, including how the different textual embeddings as features improve the results. We also visualise our results based on the T-SNE projection of the embeddings. 
	\paragraph{\textbf{Relational Space}}\label{sec:relational_space}
	As the GNN models are trained on a measurable task, we can first compare the models results, to analyse the resulting embedding of the one with better performance. In Table~\ref{table:results} we show the Area Under the ROC Curve and the Average Precision, two metrics widely used to evaluate GNNs for the link prediction task~\parencite{kipf_variational_2016, zhang_link_2018}. For the six different architectures, we are using three different ways of encoding text: The traditional TF-IDF features~\parencite{kipf_semi-supervised_2017}, the sentences embedding using Doc2Vec~\parencite{hamilton_inductive_2017}, and BERT.
	As shown, the best architecture is in every case the GCN from~\parencite{kipf_semi-supervised_2017}, closely followed by the GIN~\parencite{xu_how_2019}, and GraphSage~\parencite{hamilton_inductive_2017}. GCN is also the one that improves the most when using BERT, achieving the best performance of 0.91 in both AUC and Average Precision. Using either TF-IDF or D2V achieves approximately equal results. 
	
	\begin{table}[!t]
		\centering
		\caption{Link prediction results. Area Under the Curve and Average Precision. Mean result of 10 runs and standard deviation in parenthesis.}
		\label{table:results}
		\begin{tabular}{llll}
			\toprule
			Text Encoding     & Model     & AUC           & AP         \\   \midrule
			\multirow{6}{*}{TF-IDF}&  GAT &   0.79 (0.0) &  0.78 (0.01) \\
			& GraphUNet &  0.79 (0.03) &  0.77 (0.04) \\
			& AGNN &  0.83 (0.01) &  0.78 (0.02) \\
			& SAGE &   0.85 (0.0) &  0.87 (0.01) \\
			& GIN &  0.87 (0.01) &  0.88 (0.01) \\
			& GCN &   0.87 (0.0) &   0.89 (0.0) \\ \hline
			\multirow{6}{*}{D2V} & GAT &   0.79 (0.0) &  0.77 (0.01) \\
			& GraphUNet &  0.79 (0.03) &  0.76 (0.03) \\
			& AGNN &  0.83 (0.02) &   0.8 (0.02) \\
			& SAGE &   0.85 (0.0) &  0.87 (0.01) \\
			& GIN &  0.87 (0.01) &  0.89 (0.02) \\
			& GCN &   0.86 (0.0) &   0.88 (0.0) \\ \hline
			\multirow{6}{*}{BERT} & GAT &  0.78 (0.01) &  0.76 (0.01) \\
			& GraphUNet &  0.79 (0.03) &  0.76 (0.06) \\
			& AGNN &  0.84 (0.02) &  0.79 (0.02) \\
			& SAGE &   0.87 (0.0) &   0.89 (0.0) \\
			& GIN &  0.87 (0.02) &  0.88 (0.02) \\
			& GCN &  \textbf{0.91} (0.01) &   \textbf{0.91} (0.0) \\
			\bottomrule
		\end{tabular}
	\end{table}
	
	We performed ablation studies to understand the features relevance. In Table~\ref{table:ablation_results} we present the impact of removing one of the features from the GCN model with BERT embeddings. Results show that adding or removing the label of the first author ID does not change the performance, which corresponds with the high cardinality of this feature. Similarly, removing the organisational affiliation, subject area, topic distribution, and year only give a minor decrease in the performance. When we remove the cumulative distribution of citations the model worsens by 2\% its predictive power, and by 5\% or 6\% if we remove the BERT embedding, which means this two features are the most relevant. Not having the BERT embedding gives a closer result to those obtained using Doc2Vec and TF-IDF, which also shows the small impact of those ways of encoding text over the GNN. To conclude, the GCN is robust to the use of different features, and it is mostly focused on the network structure. We expect the embedding representation to be highly related with the network properties of the citation patterns and only mildly related with the semantic patterns through BERT. 
	
	\begin{table}[!t]
		\centering
		\caption{Link prediction results when removing a feature. Area Under the Curve and Average Precision. Average result of 10 runs and standard deviation in parenthesis.}
		\label{table:ablation_results}
		\begin{tabular}{lll}
			\toprule
			Removed &         AUC &          AP \\
			\midrule
			First Author &  0.91 (0.0) &  0.91 (0.0) \\
			Affiliation  &  0.9 (0.01) &   0.9 (0.0) \\
			Subject Area &  0.9 (0.01) &  0.9 (0.01) \\
			Topic Distribution   &  0.9 (0.01) &  0.9 (0.01) \\
			Year    &   0.9 (0.0) &   0.9 (0.0) \\
			citations at 1:10 &  0.89 (0.0) &  0.89 (0.0) \\
			BERT embedding         &  0.86 (0.0) &  0.87 (0.0) \\
			\bottomrule
		\end{tabular}
	\end{table}
	
\begin{figure}[!t]
		\centering
		\includegraphics[width=\linewidth]{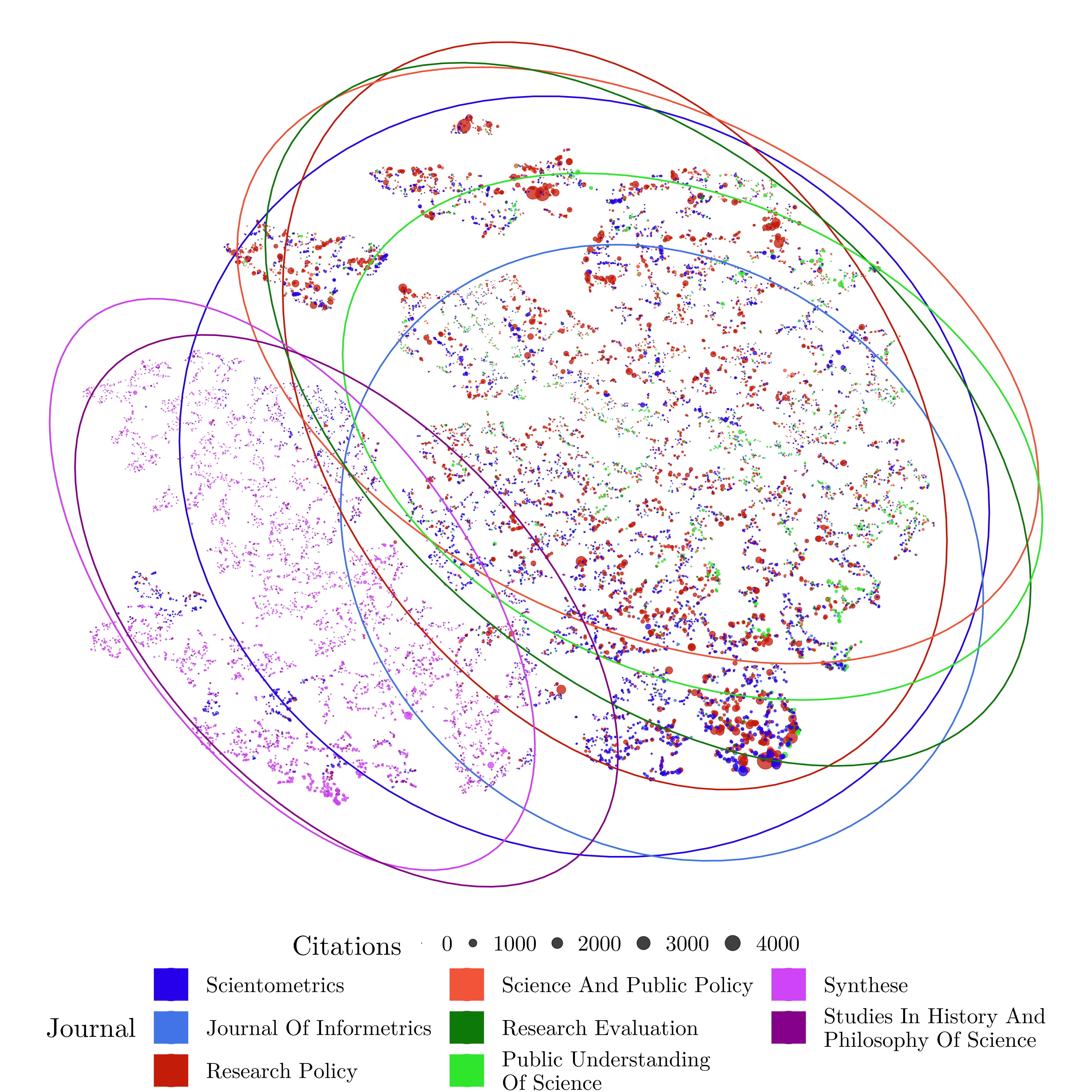}
		\caption{GNN Embedding, with GCN and BERT encoded text. T-SNE projection. Journals from the same field are presented as variations in the luminosity of a similar chromaticity. The size corresponds to the number of citations. 95\% ellipses by journal.}
		\label{fig:tsnegnnjournal}
	\end{figure}

\begin{figure}
		\centering
		\includegraphics[width=\linewidth]{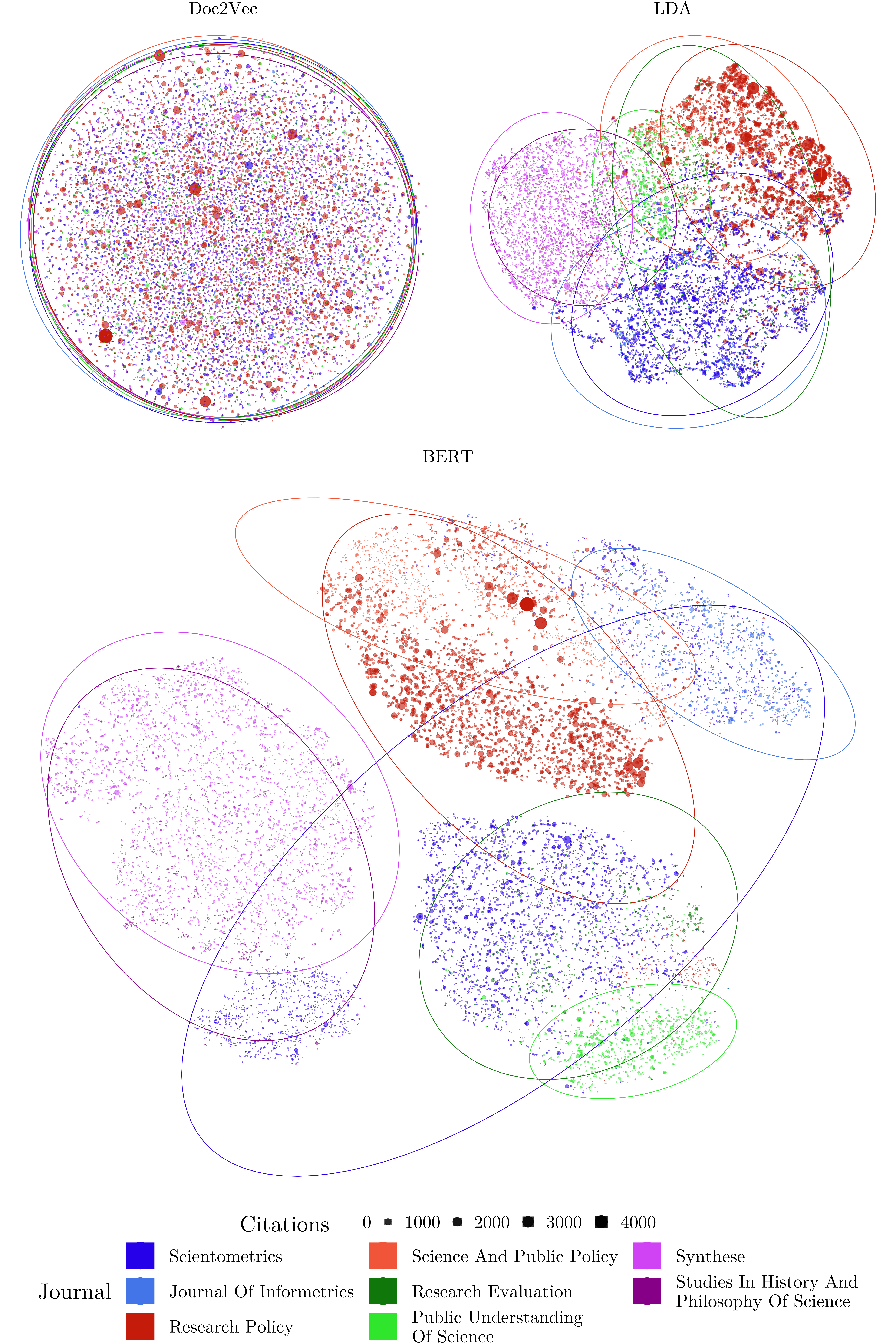}
		\caption{Semantic Embeddings. T-SNE projection. Journals from the same field are presented as variations in the luminosity of a similar chromaticity. The size corresponds to the number of citations. 95\% ellipses by journal.}
		\label{fig:semantic_embedding}
	\end{figure}

	In Figures~\ref{fig:tsnegnnjournal} and  \ref{fig:semantic_embedding}, we show the T-SNE projection \parencite{van_der_maaten_visualizing_2008}, for a sample of two journals per field, coloured by field and journal, and sized by number of citations. We can also see the 95\% ellipses of each journal, i.e., containing the 95\% of the articles of the corresponding journal~\parencite{fox2018}. Figure~\ref{fig:tsnegnnjournal} is based on the GNN embedding using GCN and BERT. Results show that there is a correlation between the number of citations and the position in space. For example, articles from the journals \textit{Research Policy} and \textit{Scientometrics} that present the highest number of citations are located in the bottom-right and top of the T-SNE projection. 
	Journals from the field of \textit{History and Philosophy of Science}, like \textit{Synthese} and \textit{Studies in History and Philosophy of Science}, where the citing culture and the selection bias of the data set (see chapter 2) imply lower citations, are located on the left of the plot. Within this field, articles with a higher number of citations cluster together in the top-left of the T-SNE representation. All fields but \textit{History and Philosophy of Science} form a uniform point cloud that correlates more with the number of citations than the corresponding journal. This can also be seen in the overlapping ellipses of most journals, except for those from \textit{History and Philosophy of Science}. The organisation of the plot by citation patterns rather than thematically highlights when compared with the semantic embeddings in Figure~\ref{fig:semantic_embedding}.  This result is in line with the expectation of the GNN, paying more attention to the relational patterns rather than the semantic content of research articles.
	Citation rates differ among disciplines. Especially research articles in the social sciences and humanities have a lower citation rate than articles in other disciplines. Further, the selection bias of the data set in favour of specific journals might cause an underestimation of citation rates for those journals, which might have a higher number of cross-references to journals that are not included in our sample as well as to other publication formats (monographs, contributions to edited volumes, etc.) that have not been analysed in this study. 
	
	\paragraph{\textbf{Semantic Space}}\label{sec:semantic_space}
	Figure~\ref{fig:semantic_embedding} shows results for the three textual embeddings.
	The Doc2Vec model is the only one of the three proposed models that was designed for document-level embeddings. Nevertheless, it does not show any correlation among journals or number of citations; the shape of the point cloud is spherical. This might be an indication that Doc2Vec is not able to learn a good representation of the documents of our data set. This is probably due to the fact that the Skip-Gram model is usually trained on millions of data points. As outlined above, pre-trained embeddings built on a general purpose corpus can improve the words representation, but this is not possible on the document level in Doc2Vec. 
	
	The LDA embedding correctly delimitates the four sub-fields of the data set. Articles from the field of \textit{History and Philosophy of Science} are located on the left side of the projection, with articles from \textit{History and Philosophy of Science} located closer to the border with the field of \textit{Other Social Sciences}. This latter field is in between \textit{History and Philosophy} on its left and the field of \textit{Management} to its right, where articles, especially from \textit{Research Evaluation}, tend to merge. We can also see a small number of articles from this field in the middle of the cluster of \textit{Library and Information Sciences}. Finally, the fields of \textit{Management} and \textit{Library and Information Sciences}, although they are well defined, share a point of contact in the centre of the plot, and some articles from each of these fields can be located in the cluster of the other. Compared with Figure~\ref{fig:tsnegnnjournal}, the correlation of highly cited articles is driven by the journals, as within each field we cannot see any clustering of highly cited articles. 
	The BERT embedding, where we were able to use pre-trained word vectors, shows a better performance than Doc2Vec, even when the model is not originally designed for document-level representation, and we are averaging the word embeddings in each document. The T-SNE representation shows a stronger delimitation between fields, and a delimitation between journals within each field. Articles from \textit{History and Philosophy of Science} are located on the left side of the figure, but with a stronger delimitation. In the top of the figure we can see the field of \textit{Management}, where articles from \textit{Research Policy} are placed towards the centre of the figure, and those from \textit{Science and Public Policy} are shifted to the top of the figure. The field of \textit{Library and Information Sciences} is split in three groups: Most of the articles from \textit{Scientometrics} lay on the centre of the figure, mixed with those from \textit{Research Evaluation}. On the left, a small proportion of the articles from this journal stays closer to the field of \textit{History and Philosophy of Science}. On the right, another group of articles from this journal, and most the articles from the \textit{Journal of Informetrics} is closer to those from \textit{Science and Public Policy}. This might be a reflection of a methodological field that develops technical aspects, but also applies the developed methodologies to thematically specific research questions. Finally, the field of \textit{Other Social Sciences} is found in the bottom of the plot, with articles from \textit{Public Understanding of Science} more delimited, and articles from \textit{Research Evaluation} closer to the field of \textit{Library and Information Sciences}.


	\subsection{Comparing the Differences between the Relational and Semantic Spaces}
	\label{sec:Differences}
	After studying the overall quality of the different models, we focus on those that have the best performance in both the semantic and the relational space. For the latter, we select the GCN using BERT embeddings as features. For the semantic space, we mainly focus on the BERT model, but also show some results for the LDA model for comparison. In this section, we show how the embedding representation of articles change in the semantic and relational space. For this, we compare the results on four different topics largely studied in the field of Science of Science: First, the representation of collaboration patterns, second, the Matthew effect in science, third, we perform a country level analysis, and fourth, the epistemic practice division in the field. Our goal is to compare how these different phenomenon are encoded in the resulting embedding, and how their representations differ within the proposed models.
	\paragraph{\textbf{Collaboration Patterns}}
	The pairwise similarity between articles can be observed through the cosine similarity between their vector. To compare the higher level groups, we can calculate the average cosine similarity between groups.
	
	One possible dimension of analysis are the collaboration patterns in terms of co-authored papers, and whether they are encoded in the embeddings. For this, we the divide articles between four groups by their forms of collaboration: A) Single author, B) internal collaborations within a single organisation, C) collaborations between authors with different organisational affiliations from the same country, and D) international collaborations, including authors from different countries and organisational affiliations. To avoid biases due to different collaboration patterns by journal, the results will be illustrated by reference to a comparison of the journals \textit{Research Policy} and \textit{Scientometrics}, i.e. comparing the average cosine similarity of articles from one journal to the other, and within each journal. Each of the three embeddings shows different uses of space, and hence for each we define a specific colour scale.
	
\begin{figure}[!t]
		\centering
		\subfloat[GNN]{\label{fig:cosine_similarity_gnn}
			\includegraphics[width=.75\linewidth]{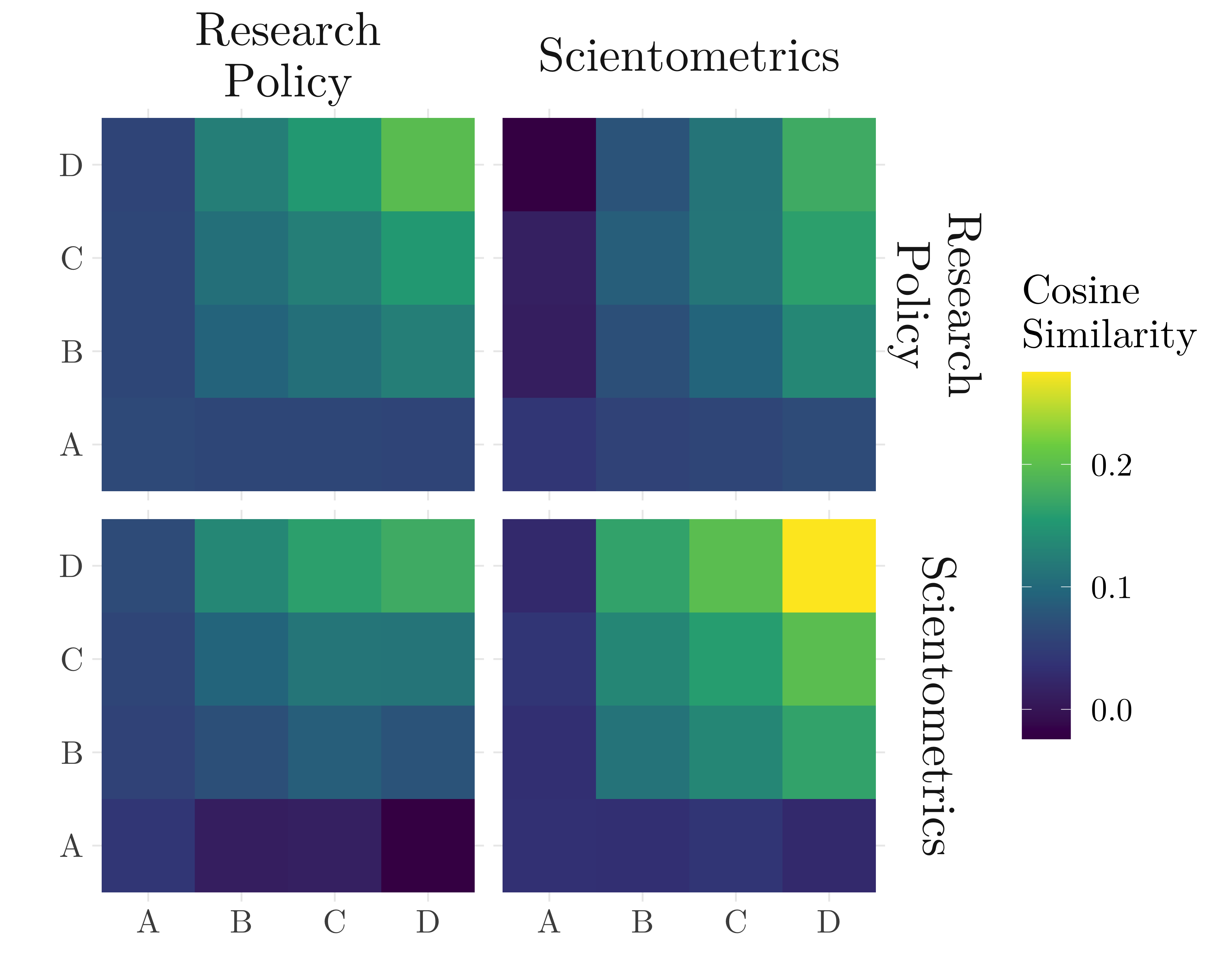}} \\
		\subfloat[LDA]{\label{fig:cosine_similarity_lda}
			\includegraphics[width=.48\linewidth]{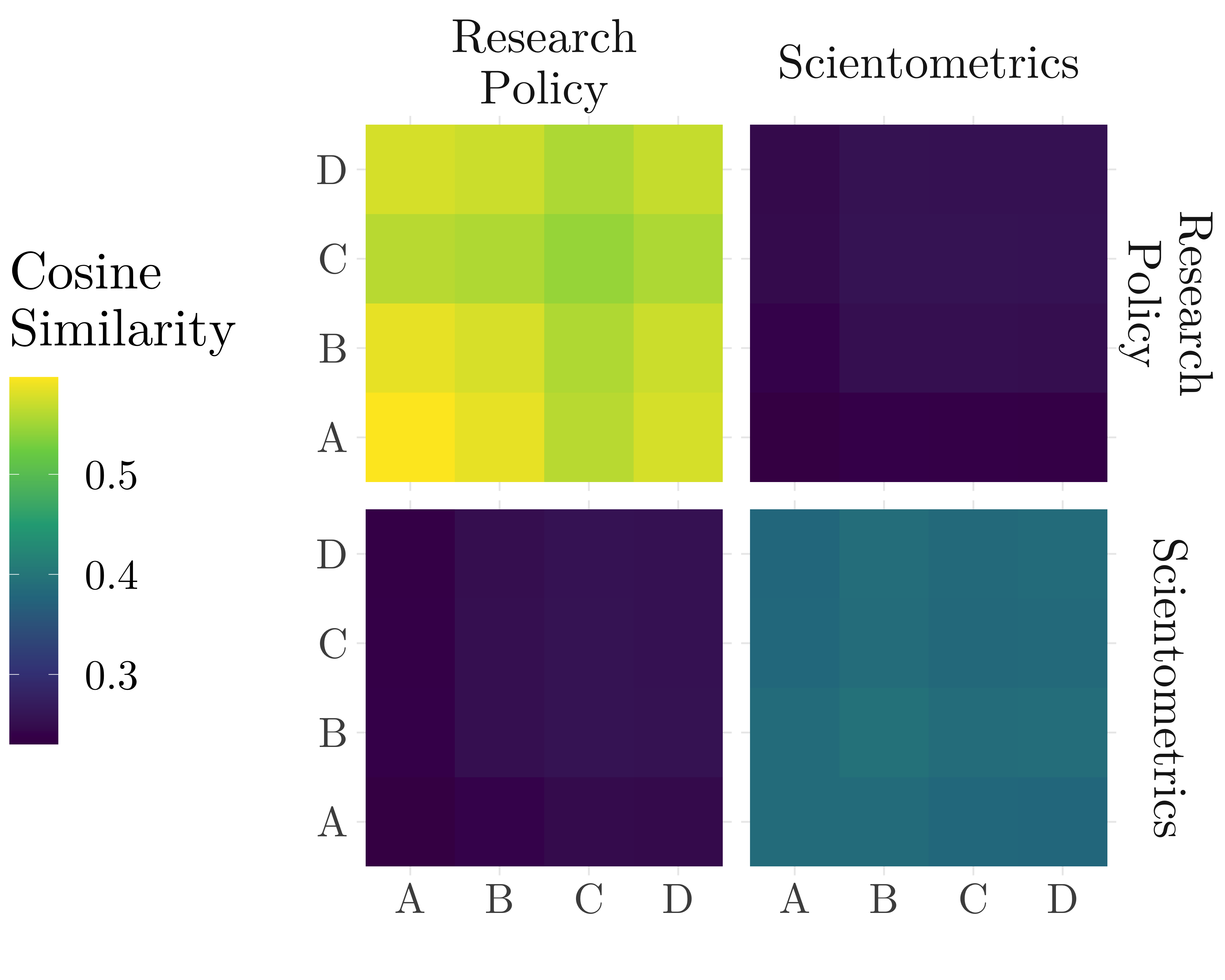}}
		\subfloat[BERT]{\label{fig:cosine_similarity_bert}
			\includegraphics[width=.48\linewidth]{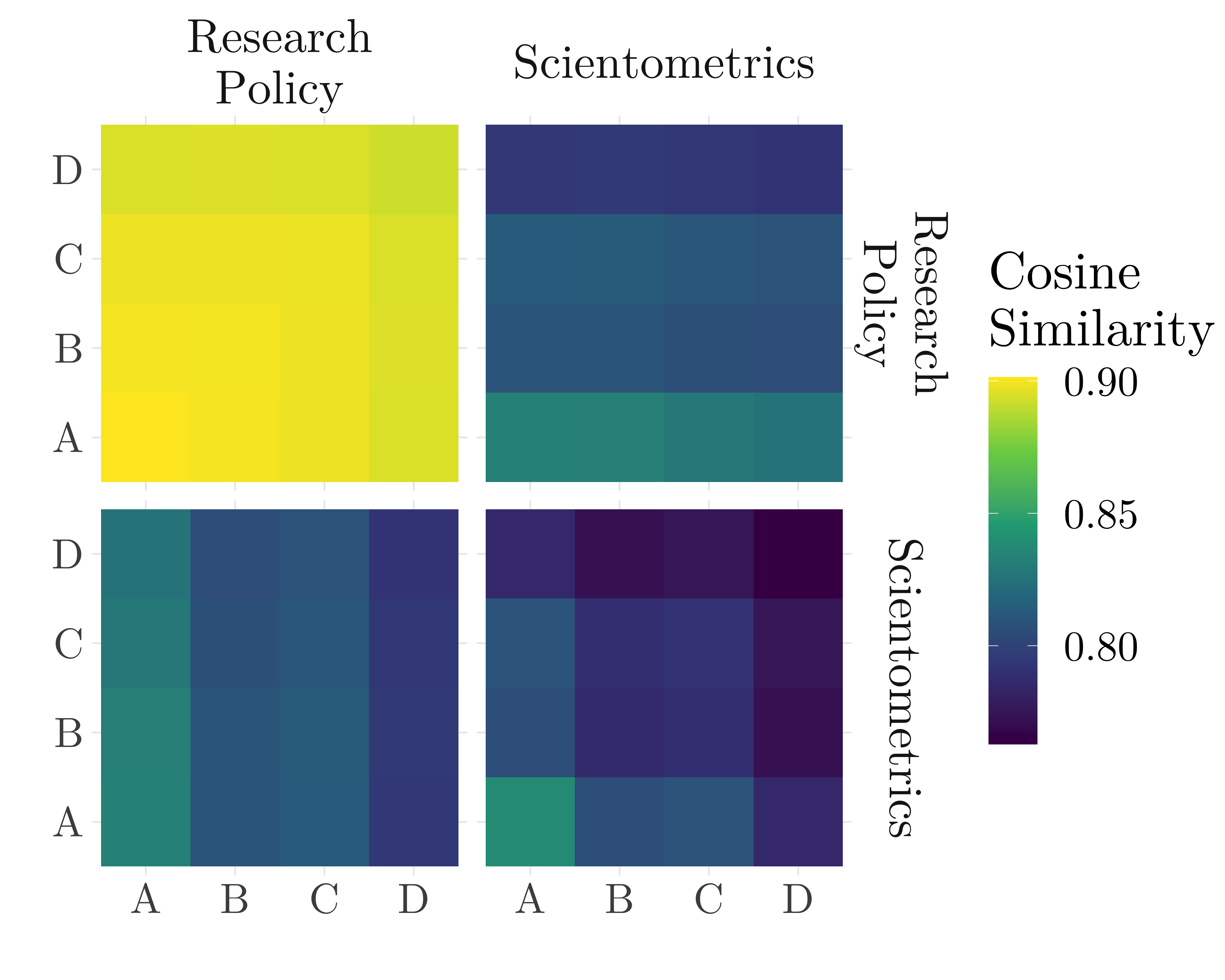}}
		\caption{Cosine Similarity by collaboration status, controlled by journal. \textbf{A}: Single authorship; \textbf{B}: Collaborations between authors from the same institution; \textbf{C}: Collaboration between authors from different institutions from the same country; \textbf{D}: Collaborations between authors from institutions in different countries.}
		\label{fig:cosine_similarity}
	\end{figure}

	Figure~\ref{fig:cosine_similarity_gnn} shows the results for GNN. Results shows a higher cosine similarity for international collaboration. Average similarity reduces from this type of collaboration toward single authorship, and this is a pattern that remains independently of the journal, i.e., articles from \textit{Research Policy} are not closer to other articles in \textit{Research Policy} than to articles in \textit{Scientometrics}, except for single author papers, where we find a small difference. This means that international collaborations are closer to all other articles. If we consider the relational space as a hyper-sphere of articles, i.e., an sphere that lays in more than three dimensions, international collaborations are in the centre of this hyper-sphere, and single authored publications tend to be on the periphery. This is inline with the bibliometric analysis on the higher impact of international collaborations measured through higher citation rates~\parencite{van_raan_influence_1998,persson_inflationary_2004, adams2013fourth}. 
	Figure~\ref{fig:cosine_similarity_lda} shows the results for the LDA model, while Figure~\ref{fig:cosine_similarity_bert} shows the results on the BERT model. In both cases, these semantic embeddings show no strong correlation between collaboration patterns and article similarity. Articles from \textit{Research Policy} are closer to each other, as well as \textit{Scientometrics} articles between them. In both cases, articles from \textit{Research Policy} tend to be closer together, and in the case of BERT, single authored articles from \textit{Scientometrics} tend to be closer to \textit{Research Policy} publications. This means that while the semantic embeddings build similar representations for articles in the same journal, the GNN builds a representation of articles where the international collaborations are similarly encoded and in a central position.

	\paragraph{\textbf{The Matthew Effect of Science}}

	The Matthew effect in science, introduced by Robert K. Merton~\parencite*{merton_sociology_1974}, states that articles that are already highly cited have a higher chance of being cited again. Figure~\ref{fig:cosine_similarity_gnn} reflects on this via the collaboration patterns, while Figure~\ref{fig:tsnegnnjournal} also shows a correlation of articles by the number of citations. To test if the embeddings are able to capture the Matthew effect, we divided the articles by their number of total citations in the Scopus data set, by the quartiles, and a separate group for those with zero citations (five groups in total). Then, we calculated the average Frobenius norm of each articles, and aggregated the results for these five groups. When studying the distribution of the Frobenius norm by citation level on the different models\footnote{See Figure~\ref{fig:frobeniousnorm} in the Appendix}, we found that while the GNN generates a higher value for the highly cited articles, the BERT, Doc2Vec and LDA models don't follow this pattern. This results mean that the GNN is systematically representing highly cited articles differently. This is expected, given that the GNN is trained on a link prediction task, and a higher Frobenius norm is associated with a higher probability of link, i.e., citation link, via the inner product decoder~(see Section~\ref{sec:Embeddings}). Nevertheless, this results show that when we design the GNN embeddings for the link prediction task, instead of trying to predict subject areas, as in \cite{kipf_semi-supervised_2017}, we are able to capture the Matthew effect in our embedding. This is an important conclusion for future research, as it show the way in which we can use the embeddings framework for studying the Matthew effect in science. Also, we found that the semantic embeddings do not encode this phenomena, and this is also important for predictive modelling in which if the Matthew effect is encoded in the embedding, this would imply reinforcing inequalities.

	\paragraph{\textbf{Country-level Analysis}}

\begin{figure}[!t]
		\centering
		\includegraphics[width=\linewidth]{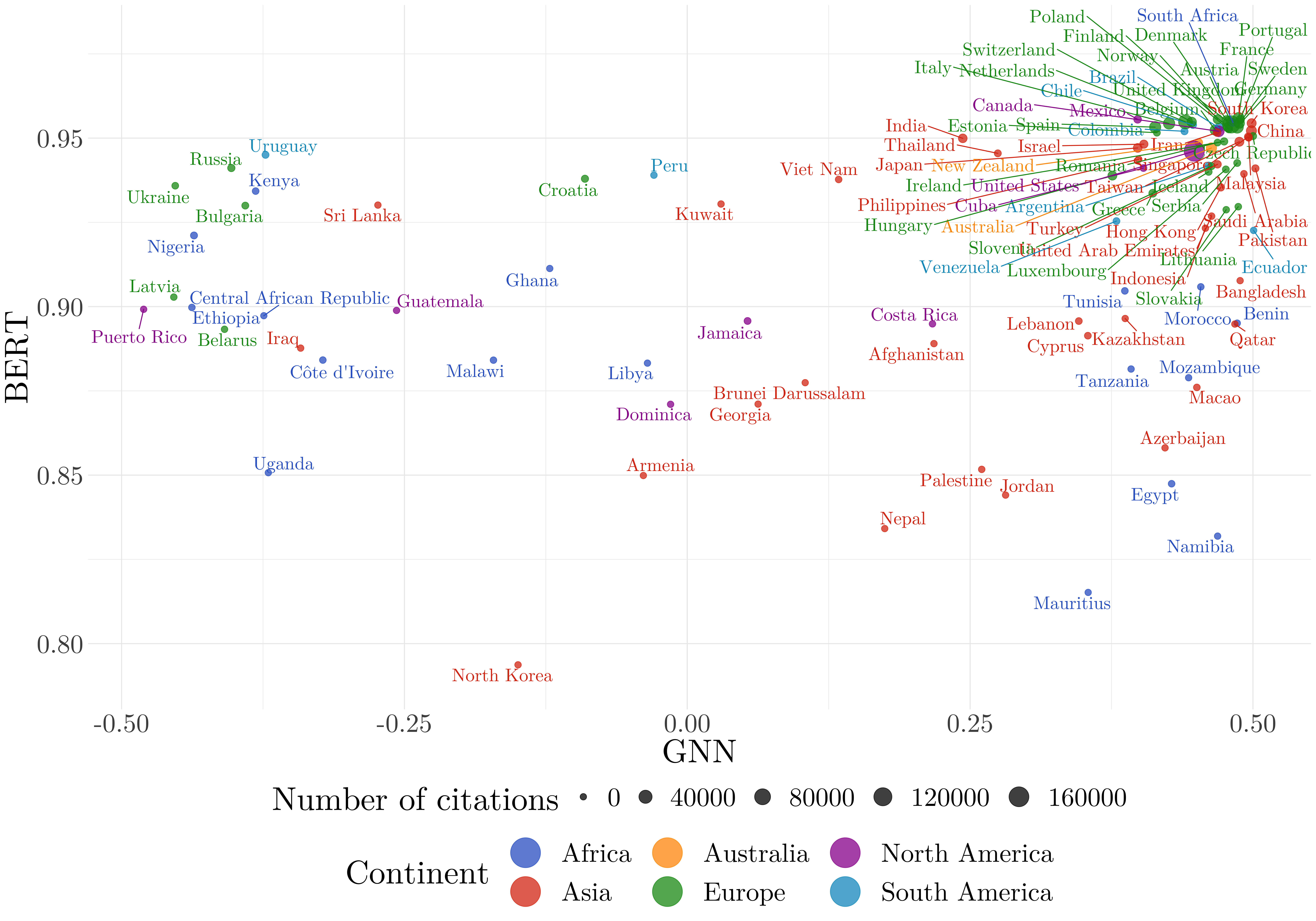}
		\caption{Average cosine similarity by countries. BERT and GNN. Size by number of citations.}
		\label{fig:average_embedding}
	\end{figure}

	In the same way we built a BERT embedding by averaging the word embeddings of each document, we can build a hierarchical representation of entities by averaging its components. One of the dimensions of analysis is the role of countries in the field of Science of Science. For this, we took the first author's organisational affiliation to ascribe a geographical location to an article. This does not necessarily mean that an article has been written in that country, but it gives us a proxy for the geographical distribution of scientific work and allows us to reconstruct the average position of countries in the embedding. Using the cosine similarity between countries, Figure~\ref{fig:average_embedding} shows the average similarity of a country with respect to all others in the GNN (horizontal axis) and BERT (vertical axis) embeddings. This means that we are comparing the semantic proximity on the vertical axis against the structural network-based proximity on the horizontal axis. Results show that there is a centre of gravity of science production~\parencite{Zhang2015} that includes most of the English-speaking countries, (Western) Europe and (East) Asia. Close to the core, we can also find some countries from (South) America and (East) Europe. South Africa is the only country from its continent close to the centre, which might be an indication for research activities of the Centre for Research on Evaluation Science and Technology (CREST) at Stellenbosch University \footnote{See \url{http://www0.sun.ac.za/crest/}}. Results also show that the BERT cosine similarity is almost always higher than 0.8, while the GNN ranges between $-0.5$ and $0.5$. This means that the semantic representation is in general very similar between all countries, while in the structural representation countries are never too close, and many of them are even in the opposite direction of most of the other countries. The presented results can be interpreted as follows: While researchers in Science of Science from all countries, within these journals, work on more or less similar content, the relevance that the academic community gives to their work is highly skewed. For example, in the case of Uruguay, the average BERT cosine similarity, i.e., based on the textual content of the articles, from this country and all others, is almost $0.95$, a really high value considering cosine similarity moves between $-1$ and $1$. On the other hand, its citation-based cosine similarity is less than $-0.35$, which means that it is in an opposite direction with respect to most of the countries. As we mentioned in Section~\ref{sec:Data}, this analysis is limited due to the limits of the data set. We cannot fully account for scientific production outside the countries that appear here as peripheral. Including journals from other regions and languages would most probably change the layout of the results, specially for the semantic embeddings~\parencite{beigel_introduction_2014}. In this sense, we have to limit the scope of interpretation to the fact that, \textit{within these journals}, the topics discussed do not vary much. Nevertheless, this result is inline with many other studies in the field on the unequal distribution of citations, at least in the international journals \parencite{demeter_world-systemic_2020, king_power_2011,bonitz_characteristics_1997, merton_sociology_1974}. 
	
	This analysis answers our first research question. If we use GAE with the GCN, we can encode the \textit{relational} dimension of articles. With the analysis of collaboration patterns, the Frobenius norm, and the country-level analysis, we can see that the idea of \textit{prestige} is captured by the GNN embeddings. This concept unfolds into different expressions, such as the different position articles have in the embedding according to their collaboration patterns and citation levels, and also on hierarchical levels of analysis, like the distribution of countries.
	
	\paragraph{\textbf{Projection of Journals on the Epistemic Spectrum}}\label{sec:quant_qual_divide}
    
    Word embeddings have shown impressive results on analogy tasks. \cite{mikolov_linguistic_2013} shows that, for word embeddings, we can solve the task ``man is to woman as king is to \rule{1cm}{0.15mm}'' by doing $\overrightarrow{\text{king}}+\overrightarrow{\text{women}}-\overrightarrow{\text{man}}$, which returns a vector close to $\overrightarrow{\text{queen}}$. This implies that there is a latent dimension of \textit{gender} in the word embedding, that can be reconstructed by the subtraction $\overrightarrow{\text{woman}}-\overrightarrow{\text{man}}$. \cite{kozlowski_geometry_2019} suggested to use analogies to describe social dimensions. \cite{kang_against_2020} applied this in the field of Science of Science. The authors define two word embeddings and compare how different concepts, like ``theory'' and ``measure'' are projected into the ``good-bad'' dimension, based on a selection of journals assigned to the quantitative and qualitative research communities.

    For our analysis, we use the analogies approach proposed by \cite{mikolov_linguistic_2013} to study the differences between journals. But instead of building different embeddings based on a pre-clasification of journals, as in~\cite{kang_against_2020}, we define a single data set and build the analogy on two pivotal journals. The ``quantitative-qualitative'' division is an open discussion~\parencite{leydesdorff_bridging_2020, weber_editors_2004} and a purely quantitative approach does not seem to be able to solve it. Given this, our analysis do not intend to prove that there is such a division. A more accurate interpretation is to simply think the analogy we propose as the \textit{latent dimension that separates epistemic practices from two journals}, being this either methodological, epistemic, ontological, or simply related to a different vocabulary typically used in different research fields.

\begin{figure}[!t]
		\centering
		\includegraphics[width=\linewidth]{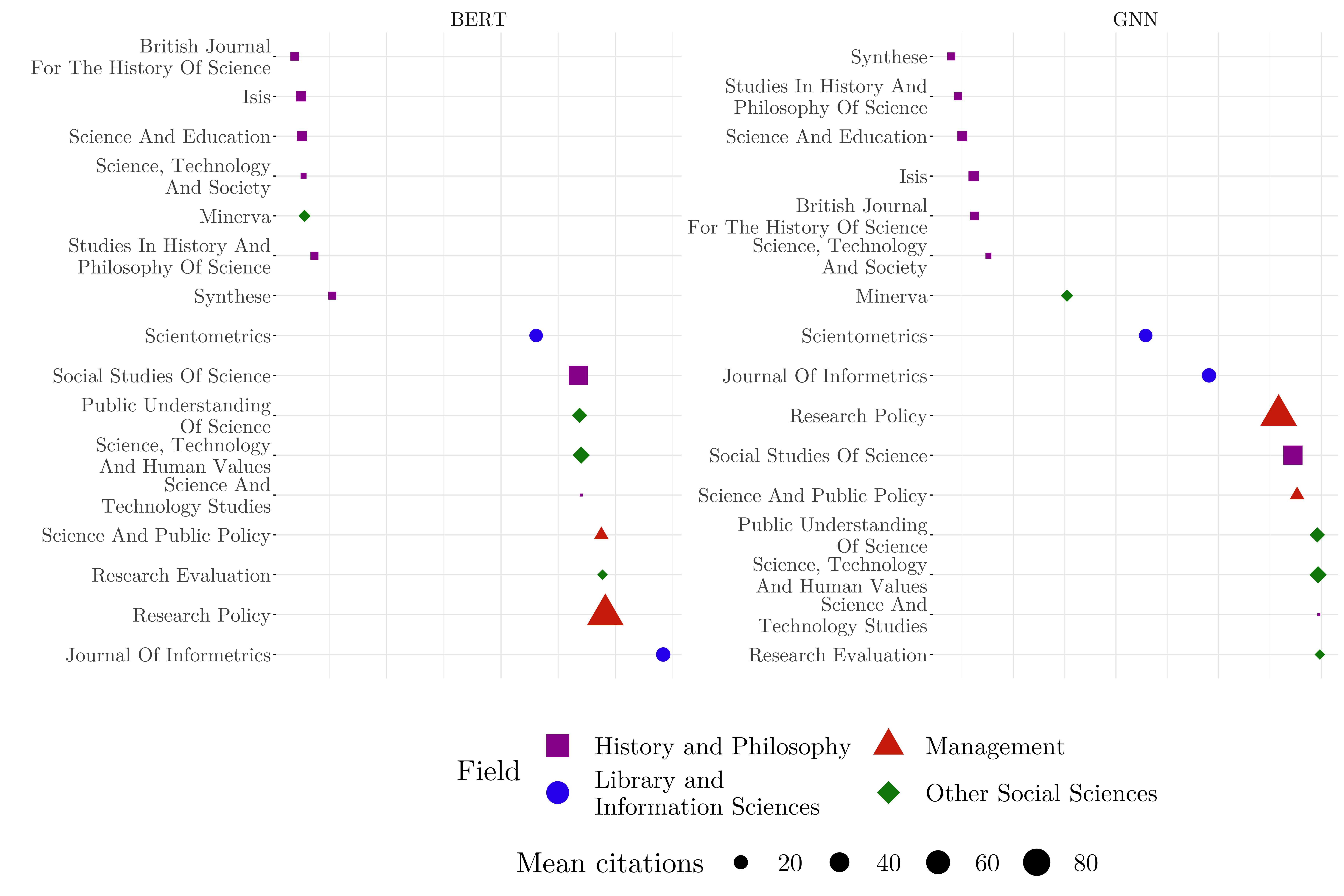}
		\caption{Cosine similarity with the \textit{Journal of Informetrics}-\textit{ISIS} dimension.}
		\label{fig:quant_qual_axis}
	\end{figure}

    For this, we first generate a vector representation of each journal in our data set, in the same way we built the representation of countries. Then, we select two journals as pivots for building the latent dimension. The selection of the pivotal journals is necessarily an arbitrary one, but compared with the approach proposed by \cite{kang_against_2020}, we do not need to previously assign each journal to one of the two poles. After this, we project the other journals to this latent dimension using cosine similarity, in this way the journals projection will be closer to one of the two poles. If the way in which journals order themselves in this dimension seems to be random, then there is no latent dimension along the two pivotal journals.
    In Figure~\ref{fig:quant_qual_axis} we consider this exercise using the journals \textit{ISIS} and \textit{Journal of Informetrics} as pivots. This latter defines its scope on ``research on quantitative aspects of information science"\footnote{See \url{https://www.journals.elsevier.com/journal-of-informetrics}.}. \textit{ISIS} is a long standing journal on ``history of science, medicine, and technology and their cultural influences''\footnote{See \url{https://www.journals.uchicago.edu/journals/isis/about}.}. In the BERT embedding we can see that the \textit{Journal of Informetrics} is on one side of the extremes, but on the other side we have \textit{British Journal for the History of Science}. We can also see that there is a strong division of journals along the axis, with seven journals really close to the \textit{ISIS} pole and nine close to the \textit{Journal of Informetrics} pole. Almost all journals in the \textit{ISIS} pole are from \textit{History and Philosophy of Science}, except for \textit{Minerva}, a multidisciplinary journal. All journals from \textit{Management} and \textit{Library and Information Sciences} are on the \textit{Journal of Informetrics} pole, although the journal \textit{Scientometrics} is not as close to the extreme as would be expected. Except for \textit{Minerva} all other journals from \textit{Other Social Sciences} are also on the \textit{Journal of Informetrics} pole. On the other hand, the same exercise using the GNN embedding gives a different result. In this case, both journals from the field of \textit{Library and Information Sciences} are together in the middle range, indicating that the \textit{Journal of Informetrics}-\textit{ISIS} dimension is poorly defined on this embedding, indicating that the GNN embedding is not capturing the semantic information as well as the BERT embedding. If we consider the mean citations by journal in the GNN embedding, there seem to be more correlation with this, than that with the epistemic differences by journal. Given this, we can conclude that while the BERT embedding can correctly capture this phenomenon, the GNN embedding is driven by the relational structure of the citation network, rather than the epistemic content of articles, and therefore is not a proper tool for this type of analysis.
    
    
    
	The outlined results answer research question two: Using text-based embeddings, we can encode a \textit{semantic} representation at the article level. Using analogies between journals, we can reflect the epistemic difference between them and project other journals, or articles onto that spectrum. 


	\section{Conclusions}
	\label{sec:Conclusions}
	In this paper, we explored the use of \textit{embeddings} as representations of research articles. We presented an overview of techniques designed on the different elements that compose an article: its text, metadata, and citations. The objective of this article is to study the use of new methodologies that are currently being developed in the field of computer science to analyse journals and research articles in the field of Science of Science. 

	In Section~\ref{sec:Embeddings}, we presented two approaches for building an embedding space of articles: the NLP techniques, and the GNN techniques. In Section~\ref{sec:semantic_space}, we found that, using textual content in document embeddings, we are able to build a semantic space. In Section~\ref{sec:relational_space}, we evaluated the performance of the different models, and we concluded that, when we use the network structure in a GNN, we define a relational space that embed the social relations underneath. We also presented an extensive comparative analysis within each group of models to find the best performing architecture and the set of hyper-parameters. Our results show that for the semantic embedding, the BERT model gives the most clear results, while for the link prediction task on the direct citation network, the GAE with GCN, using the BERT embedding along with the metadata features gives the best performance. In Section~\ref{sec:Differences}, we compared the semantic and relational spaces along four different dimensions: collaboration patterns, the country-level analysis, the Mathew effect, and the journals epistemic practice division. From this analysis, we found that the relational space captures the difference on collaboration status, citation levels, and aggregated levels of analysis like differences between countries. The semantic space on its hand, captures phenomena related with articles topics, their corresponding journal, and epistemological aspects, like the journals epistemic practice division.
	
	These are promising results as it opens the possibility for many different in-depth analysis using this different techniques, according to the respective research question. A responsible use of embeddings as tools for analysis does not imply that we should not use those embeddings that reproduce the existing bias, but that we should acknowledge it. For example, a recommendation system based on GNN would reproduce the biased attention researchers give to different articles. On the other hand, using GNN can be of great help to better understand the biases themselves. These biases, as seen on the country-level analysis, might not only be a reflection of the scientific community, but are likely related to the chosen Scopus database and the sampling of journals and articles itself.
	This paper is based on a small data set in the field of Science of Science, and therefore represents a case study. More research on other fields and cross-disciplinary data sets have to be made to explore if the present results are also valid in other fields of research as.
	The main contributions of this paper are fourfold:
	

	\begin{enumerate}
		\item \textbf{Data}:
		We have built a data set with 16 core journals from the field of Science of Science, from a range of disciplines. The data set includes 22,151 research articles, including their metadata (abstract, title, keywords, authors, organisational affiliations of the authors, year of publication, and journals subject areas), the results of the LDA model (the topics distribution of each article), and the cumulative citations from the network. From this data set, we build a network of direct citations with 16,578 nodes and 68,797 links. This data set is suitable for models that work with metadata features, NLP, and networks, allowing us to compare the performance of different approaches\footnote{The DOI of the articles that compose the data set will be available, together with the code for implementing the models in this article, at \href{https://github.com/DiegoKoz/scisci}{https://github.com/DiegoKoz/scisci} upon publication, and the full data set is available upon request and approval of Scopus. Sharing the code and data aims to improve the reproducibility of this work.}. 
		\item \textbf{Semantic Modelling}: We performed three different approaches for sentences embedding of titles, abstracts, and keywords: Doc2Vec, LDA and BERT, and compared their performance as features for a link prediction model. We presented the Topic Model results as an interactive plot, which give an overview of the topics discussed within the data set. We came to the conclusion that for a data set of these characteristics it is more powerful to use a BERT model, because it can benefit from a pre-training on a large corpus.
		\item \textbf{Relational Modelling}: We trained the GNN on the link prediction task and made an extensive comparison of multiple possible encoders, using GNN layers that are currently the state of the art in computer sciences. We show that the GCN is the best performing architecture for this task in the present context. We also performed ablation studies to find that the BERT encoding of text and the cumulative citations are the most relevant features for the model.
		\item \textbf{Models Comparison}: We compared the latent information in the different types of embeddings and arrived to the conclusion that the textual embedding generates a semantic space, while the GNN generates a relational space. This is an important methodological conclusion, as its a distinction of which type of modelling should be used given the research questions. 
	\end{enumerate}

	Multiple recommendations for future research rise from this article: First, methodological research needs to be done over other GNN architectures. In this article, we presented GNN models for the link prediction task, but node prediction tasks are also suitable for this network. Predicting the article's journal, for example, would generate an embedding representation much closer to the semantic space. Predicting the article's author with GNN can be an important improvement for the author-name disambiguation problem \parencite{schulz_exploiting_2014}. Besides, direct citation networks are not the only network structure to be consider. Second, the network that emerges from bibliographic coupling and co-citation can be compared with the results from direct citations. Third, the study of co-authorships and mobility networks are promising lines of research, in order to explore how communities of co-authors and institutions are organised in the embeddings. Fourth, the use of this methodology can generate important new insights for the field of Science of Science: the flexibility of the low-level representations allows a quantitative approach to many research questions. Problems like the Matthew or the Mathilda effects~\parencite{rossiter_matthew_1993} can be approached using the cosine similarity analysis presented in this article. The time dynamics in this phenomena can also be studied by simply splitting the data set in decades and building multiple embeddings, as in \cite{Garg2018}. Also, as we have seen in this article, the field classification on journal level can be a problematic task. Embeddings methodologies can be used for field classification on article level. Finally, the development of methodologies based on relational embeddings enable researchers to detect gender and ethnic biases also carries the potential use in policy recommendations.

	\section*{Acknowledgments}
	
	\noindent
	{\it Funding}.
	This work has been supported by the Doctoral Training Unit \textit{Data-driven computational modelling and applications} (DRIVEN, \url{https://driven.uni.lu}), which is funded by the Luxembourg National Research Fund under the PRIDE programme (PRIDE17/12252781).
	
	{\it Acknowledgments}.
	The authors would like to thank Cassidy R. Sugimoto and Vincent Larivière for their helpful input on the selection of journals for the data set. We would also like to thank Justin J.W. Powell, Marcelo Marques, and Mike Zapp for their valuable comments on previous versions of the manuscript.

\pagebreak
	
	\printbibliography
	
	\pagebreak	
	\begin{appendices}
		
		\renewcommand{\thetable}{\thesection\arabic{table}}
		\renewcommand\thefigure{\thesection.\arabic{figure}}    
		
		\section{Glossary} \label{sec:glossary}
		
		\setcounter{table}{0}
		\setcounter{figure}{0}

		\begin{table}[htbp]
			\centering
			\caption{Glossary of acronyms.}
			\label{table:glossary} 

			\captionsetup{justification=centering}
			\begin{tabular}{@{}ll@{}}
				\toprule
				Meaning                                                & Acronym \\ \midrule
				Natural Language Processing                            & NLP     \\
				Document-Term Matrix                                   & DTM     \\
				Graph Neural Networks                                  & GNN     \\
				Term Frequency - Inverse Document Frequency            & TF-IDF  \\
				Latent Dirichlet Allocation                            & LDA     \\
				Recurrent Neural Network                               & RNN     \\
				Convolutional Neural Networks                          & CNN     \\
				Convolutional Graph Networks                           & CGN     \\
				Graph Convolutional Networks                           & GCN     \\
				Graph Isomorphic Network                               & GIN     \\
				Graph Attention Network                                & GAT     \\
				Attention-based Graph Neural Networks                  & AGNN    \\
				Graph Autoencoder                                      & GAE     \\
				Area Under the Receiver Operating Characteristic Curve & AUC     \\
				Average Precision                                      & AP      \\
				True Positive Rate                                     & TPR     \\
				False Positive Rate                                    & FPR     \\ \bottomrule
			\end{tabular}
		\end{table}
		
		\pagebreak

		\section{Data statistics. In and Out of Network}\label{sec:oon_stats}
		\setcounter{table}{0}
		\setcounter{figure}{0}    
		
		In this section, we split the table \ref{table:network_stats} between those articles that are part of the network in Table \ref{table:in_the_net}, and those that have no information about references and are not referenced by any other article, in Table \ref{table:out_of_net}. We can see that 75\% of articles are part of the network, and compared to those out of the network, they have a higher mean citation. Most of the articles that cannot be included in the network are from `Synthese' (30\%), `Research Policy'(12\%), or `Science and Public Policy' (11\%).

		\begin{table}[htbp]
			\centering
			\caption{Articles out of the network summary statistics.}
			\label{table:out_of_net}

			\resizebox{\textwidth}{!}{%
				\begin{tabular}{llrrr}
					\toprule
					Field & Journal & \begin{tabular}[c]{@{}r@{}}Articles\\ Retrieved\end{tabular} & \begin{tabular}[c]{@{}r@{}}Mean\\ Citations\end{tabular} & \begin{tabular}[c]{@{}r@{}}Max\\ Citations\end{tabular} \\ \midrule
					\multirow{2}{*}{Management}                                                                             & Research Policy                                                                         & 643                                                         & 79.72                                                    & 3404                                                    \\
					& Science And Public Policy                                                               & 637                                                         & 10.31                                                    & 409                                                     \\ \hline
					\multirow{2}{*}{\begin{tabular}[c]{@{}l@{}}Library and \\ Information Sciences\end{tabular}}            & Scientometrics                                                                          & 784                                                         & 19.73                                                    & 435                                                     \\
					& Journal Of Informetrics & 12                                                          & 13.50                                                    & 28                                                      \\ \hline
					\multirow{8}{*}{History and Philosophy of Science}  & Synthese                                                                                & 1702                                                        & 6.70                                                     & 564                                                     \\
					& \begin{tabular}[c]{@{}l@{}}Studies In History And \\ Philosophy Of Science\end{tabular} & 372                                                         & 7.68                                                     & 63                                                      \\
					& Isis & 250                                                         & 9.09                                                     & 123                                                     \\
					& Science And Education & 250                                                         & 10.92                                                    & 298                                                     \\
					& \begin{tabular}[c]{@{}l@{}}British Journal For\\ The History Of Science\end{tabular}    & 145                                                         & 9.03                                                     & 54                                                      \\
					& Social Studies Of Science & 139                                                         & 24.71                                                    & 648                                                     \\
					& Science, Technology And Society & 134                                                         & 4.15                                                     & 47                                                      \\
					& Science And Technology Studies   & 8                                                           & 2.25                                                     & 8                                                       \\ \hline
					\multirow{5}{*}{\begin{tabular}[c]{@{}l@{}}Education, \\ Communication and\\ Anthropology\end{tabular}} & Public Understanding Of Science                                                         & 170                                                         & 25.32                                                    & 416                                                     \\
					& Research Evaluation                                                                     & 142                                                         & 6.15                                                     & 76                                                      \\
					& \begin{tabular}[c]{@{}l@{}}Science, Technology\\ And Human Values\end{tabular}          & 106                                                         & 19.51                                                    & 272                                                     \\
					& Minerva & 79 & 7.24 & 78 \\ \hline
					& Total & 5573 & 16.00 & 3404 \\ \bottomrule 
				\end{tabular}
			}
		\end{table}
		
		\begin{table}[htbp]
			\centering
			\caption{Articles in the network summary statistics.}
			\label{table:in_the_net}

			\resizebox{\textwidth}{!}{%
				\begin{tabular}{llrrr}
					\toprule
					Field & Journal & \begin{tabular}[c]{@{}r@{}}Articles\\ Retrieved\end{tabular} & \begin{tabular}[c]{@{}r@{}}Mean\\ Citations\end{tabular} & \begin{tabular}[c]{@{}r@{}}Max\\ Citations\end{tabular} \\ \midrule
					\multirow{2}{*}{Management} & Research Policy & 2578 & 84.75 & 4820 \\
					& Science And Public Policy & 1070 & 15.03 & 462 \\ \hline
					\multirow{2}{*}{\begin{tabular}[c]{@{}l@{}}Library and\\ Information Sciences\end{tabular}} & Scientometrics & 4352 & 20.10 & 1334 \\
					& Journal Of Informetrics & 864 & 22.76 & 352 \\ \hline
					\multirow{8}{*}{History and Philosophy of Science} & Synthese & 2449 & 9.80 & 910 \\
					& Social Studies Of Science & 930 & 43.38 & 4709 \\
					& Science And Education & 784 & 11.82 & 177 \\
					& \begin{tabular}[c]{@{}l@{}}Studies In History\\ And Philosophy Of Science\end{tabular} & 539 & 9.50 & 145 \\
					& Isis & 273 & 15.57 & 415 \\
					& Science, Technology And Society & 211 & 7.28 & 122 \\
					& \begin{tabular}[c]{@{}l@{}}British Journal\\ For The History Of Science\end{tabular} & 131 & 10.16 & 88 \\
					& Science And Technology Studies & 103 & 5.52 & 39 \\ \hline
					\multirow{5}{*}{\begin{tabular}[c]{@{}l@{}}Other Social Sciences:\\ Education, Communication\\ and Anthropology\end{tabular}} & Public Understanding Of Science & 807 & 26.04 & 518 \\
					& \begin{tabular}[c]{@{}l@{}}Science, Technology\\ And Human Values\end{tabular} & 651 & 35.04 & 828 \\
					& Research Evaluation & 524 & 15.04 & 223 \\
					& Minerva & 312 & 18.86 & 624 \\ \hline
					& Total & 16578 & 21.92 & 4820 \\ \bottomrule 
				\end{tabular}
			}
		\end{table}
		
		\pagebreak
		
		\section{Models definitions}
		\setcounter{table}{0}
		\setcounter{figure}{0}    
		
		\subsection{LDA}
		\label{sec:lda_def}
		As a generative Bayesian model, LDA states a generative process in which data is created, and then uses Bayes theorem to fit the parameters. The generative process is as follows:
		
		\begin{itemize}
			\item Each topic is generated as a multinomial distribution, $\beta_i$, over words
			\item For each document in the corpus, the words are defined in a two-step process:
			\begin{enumerate}
				\item First, we define the distribution of topics in the document as a multinomial distribution, $z$,
				\item for each word:
				\begin{itemize}
					\item randomly select a topic from the given $z$ distribution,
					\item given the topic, randomly select the word from the corresponding $\beta$ distribution.
				\end{itemize}
			\end{enumerate}
		\end{itemize}
		
		If we have a dictionary of $V$ possible words, $n$ documents, and $k$ topics. $\beta$ will be a matrix of $k \times V$, where each row is the realisation of a Dirichlet process, i.e., a stochastic process where its realisations are multinomial distributions. $\beta$ will indicate the distribution of words over topics. $Z$ will be a matrix of $n \times k$ which will indicate the distribution of topics over documents. $\beta$ and $Z$ are the desired outputs, but the corpus only gives us the actual words of documents. If we consider these documents as a realisation of this chained random processes, we can use the Bayes Theorem to infer the probabilistic distributions: 
		$$
		p(\theta,z|w,\alpha,\beta) = \frac{p(\theta,z,w|\alpha,\beta)}{p(w|\alpha,\beta)}
		$$
		where $\theta$ is the Dirichlet process that defines the distribution of topics over documents, $z$, and $\alpha$ is its parameter.
		
		\subsection{GNN Models}
		\label{sec:gnn_models}
\begin{figure}[htbp]
			\centering
			\includegraphics[width=.75\linewidth]{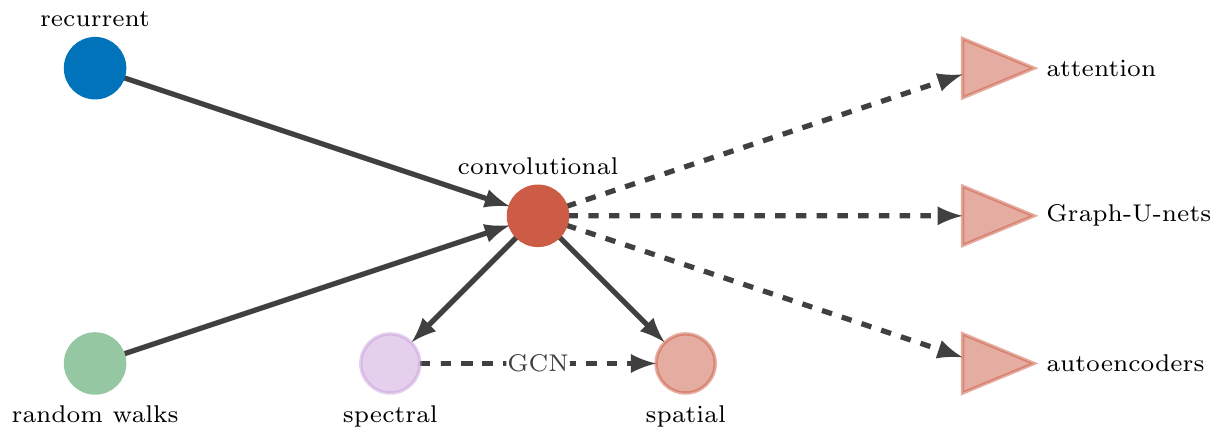} 	
			\caption{Graph Neural Networks framework.}
			\label{fig:timeline}
		\end{figure}

		Figure \ref{fig:timeline} shows a synthesis of the different approaches taken to solve this issue. The first approaches were based on building sequences using random walks \cite{perozzi_deepwalk_2014}, or recurrent models \cite{scarselli_graph_2009}. After this, Graph Convolutions were defined using spectral methods \cite{bruna_spectral_2014} and spatial methods \cite{hamilton_inductive_2017}. More recently, new architectures were proposed that incorporate attention mechanisms \cite{velickovic_graph_2018}, U-nets \cite{gao_graph_2019} and autoencoders \cite{kipf_variational_2016}. For the remaining of this section, we present these models' intuitions, for an in-depth literature review, we refer the readers to \cite{hamilton_representation_2017,bacciu_gentle_2020, zhou_graph_2018,wu_comprehensive_2020}.
		
		\subsubsection{First Approaches}
		In this section, we briefly present the first approaches for GNN. These models are not used in the subsequent experiments, as they are not currently the state of the art. Nevertheless, they are all conceptually important.
		
		For the task of \textit{node embedding}, given the developments on Word Embeddings \cite{mikolov_linguistic_2013}, one of the first strategies proposed was to use random walks over nodes to define a sequence that can be later used as the input for a Word2Vec model, as it is normally done with texts on NLP. The first model that proposed this technique was DeepWalk \cite{perozzi_deepwalk_2014}. Later, \cite{grover_node2vec_2016} proposed node2vec, which defines flexible biased random walks, that includes parameters for adjusting the path taken by the random walks to search for structural roles or community structures. The major problem with these approaches is that they do not consider the features of the nodes, so they miss potentially useful information.
		
		\cite{scarselli_graph_2009} proposed the Graph Neural Network model which iteratively updates the nodes' state looking at its neighbours, until it converges. This recurrent model uses a single layer  which is iteratively updated. 
		
		Convolutional Graph Networks (CGN) models instead use a stack of layers. In this way, the number of updates is fixed, and the parameters of each layer are allowed to change, giving more flexibility to the model. 
		Spectral-based methods were the first type of CGN \cite{bruna_spectral_2014}. They use the \textit{graph Fourier transform} on the Laplacian matrix (a normalised adjacency matrix), which can be thought of as the effect of a signal over the network. This model, while conceptually important, suffers from many problems. In particular, as it uses the full graph structure, it can only work on transductive settings, i.e., it cannot be used on a different graph on train and test. More importantly, the eigen-decomposition requires $\mathcal{O}(n^3)$ where $n$ is the number of nodes. This is prohibitively expensive when the network has billions of nodes, like social networks.
		
		\subsubsection{State Of The Art}
		In this section, we present the current state of the art in GNN.  A combination of these models will be used in the experimental analysis for the task of link prediction. 
		
		\paragraph{GCN}
		
		Many models improve the limitations present in the spectral method proposed by \cite{bruna_spectral_2014}. \cite{defferrard_convolutional_2016} build an approximation of the original model with Chebyshev polynomials. \cite{kipf_semi-supervised_2017} introduce the \textit{Graph Convolutional Networks} (GCN), which further reduces the model and includes self-connections, this means that in the iterative process of building a representation based on its neighbours, the node will also look at itself, which is a desirable property.  The GCN simplifies the model by only looking at the first-order neighbourhood. If the representation of a node is initiated with its feature vector, the GCN update builds an average representation based on itself, due to self-connections, and its neighbours. Instead of iterating the update step until convergence, like recurrent models, in GCN a stack of layers is built. The stacking of GCN layers allows the  node representation to be based on more distant nodes. Here, simplicity is the key for building a powerful model. 
		
		\paragraph{GraphSage}
		
		\cite{hamilton_inductive_2017} propose several variations over the GCN, in the  \textit{GraphSage} model. As the formulation of the problem can be useful for understanding how Graph Convolutional Networks works, we present their algorithm in \ref{alg:graphsage} . The model needs the following inputs:
		\begin{itemize}
			\item The Graph $\mathcal{G}(\mathcal{V},\mathcal{E})$ with a list of vertices $\mathcal{V}$ and edges $\mathcal{E}$, and
			\item the input features $x$, where $x_v$ is the feature vector of the node $v$.
		\end{itemize}
		We also need to define the number of layers the model will have, $K$, a set of weight matrices $W^k$ for each layer (that will be later trained with the data), and an activation function $\sigma$. We also need a way in which a group node embeddings will be aggregate, and a way of defining the neighbourhood of a node.  
		
		The model starts the node embeddings with their feature vectors. After this, in each of the $K$ layers, for each node $v \in \mathcal{V}$, it first defines its neighbours. One of the changes with respect to GCN is that GraphSage samples a fixed amount of neighbours to control the computational footprint. Given the neighbourhood, the $AGGREGATE$ function is used to build a new vectorised representation of those (line 4), and this is later concatenated with the embedding of node $v$ in the previous layer (line 5). A projection is made with the $W^k$ matrix, and also an activation layer is used. This correspond with the typical structure of a deep learning layer. When the model is fitted with back-propagation for a specific task, the $W^k$ are updated in order to optimise the lost function \cite{kelley_gradient_1960}. Line 6 is simply a regularisation.
		
		\begin{algorithm}[htbp]
			\caption{GraphSAGE \cite{hamilton_inductive_2017}}
			\label{alg:graphsage}
			\hspace*{\algorithmicindent{  \textbf{Input:} Graph $\mathcal{G}(\mathcal{V},\mathcal{E})$: input features $x$ ; depth K ; weight matrices $W^k, \forall k \in {1,...,K}$; nonlinearity $\sigma$ ; differentiable aggregator functions $AGGREGATE_k, \forall k \in {1,...,K}$; neighbourhood function $\mathcal{N}: v \to 2^\mathcal{V}$.}}\\
			\hspace*{\algorithmicindent}  \textbf{output}  Vector representations $z_v \forall v \in \mathcal{V}$.
			\begin{algorithmic}[1]
				
				\State $h_v^0 \leftarrow x_v, \forall v \in \mathcal{V}$;
				\For {$k= 1\dots K$}
				\For {$v \in \mathcal{V}$}
				\State $h_{\mathcal{N}(v)}^k \leftarrow AGGREGATE_k(\{h_u^{k-1},\forall u \in \mathcal{N}(v)\})$;
				\State $h_v^k \leftarrow \sigma(W^k CONCAT(h_v^{k-1}, h_{\mathcal{N}(v)}^k))$
				\EndFor
				\State  $h_v^k \leftarrow h_v^k/||h_v^k||_2, \forall v \in \mathcal{V}$
				\EndFor
				\State $z_v \leftarrow h_v^K, \forall v \in \mathcal{V}$		
			\end{algorithmic}
		\end{algorithm}
		
		Depending on the election of the AGGREGATE and the CONCAT operators, this model is an approximation of GCN from \cite{kipf_semi-supervised_2017}. If we use the \textit{mean} as an aggregation function, and instead of concatenating $h_v^{k-1}$ and $h_{\mathcal{N}(v)}^k$ we average them, this is exactly the GCN model. The authors also propose two other aggregators: the \textit{LSTM aggregator} which can be more expressive, although the LSTM model (a RNN) requires a sequential order in the inputs, so the authors need to define an ad hoc order for the neighbours. And the \textit{pooling aggregator} in which the neighbours vector representation is fed through a fully connected layer and then the max-pooling is applied.
		
		\paragraph{GIN}
		Recent studies analyse the relation between GNN and the Weisfeiler-Lehman test of isomorphism \parencite{xu_how_2019}. Two graphs are isomorphic if they are topologically identical. Besides the embedding representation of nodes, it is also possible to build embedding representations of the entire graph. In the previous framework, we would need to add a READOUT operation that takes the nodes embedding as input and generates a single embedding representation for the entire network. If for two graphs $G_1$ and $G_2$ that are non-isomorphic we can build a different embedding representation, we are able to distinguish between those two. \cite{xu_how_2019} proves that the GNN power for discriminating between non-isomorphic networks is at most that of the Weisfeiler-Lehman test. However not every GNN can have this power, the \textit{Graph Isomorphic Network} (GIN) is a proposal that achieves the maximum discriminative capacity by using a specific update function, which replaces the $AGGREGATE$ functions proposed by \cite{hamilton_inductive_2017} (\textit{mean}, \textit{LSTM} or \textit{max}) with a \textit{summation} over the neighbourhood. This model has the advantage of a theoretically robust decision on the AGGREGATE function. 
		
		\paragraph{GAT}
		In GCN, nodes embeddings are updated using their neighbours embedding. Up to this point, all models consider that every neighbour has the same influence, which might not be true. \textit{Graph Attention Networks } (GAT), introduced by \cite{velickovic_graph_2018}, use attention mechanisms (which are currently state of the art in other problems, like NLP) to assign a different influence to each neighbour. The update of the node representation becomes:
		$$
		h_v = \sigma \left(\sum_{u \in \mathcal{N}(v)} \alpha_{u,v}Wh_u\right)
		$$
		where $W$ is still a learning weight matrix and $\alpha_{u,v}$ is the normalised attention between nodes $u$ and $v$.
		
		Following \cite{vaswani_attention_2017}, GAT uses \textit{multihead attention}, which implies applying $K$ independent attentions and concatenating their results (except for the final layer, where they are averaged).
		
		This model has the advantage of learning the different importance of the neighbours, given their feature vector. Moreover, the attention mechanisms have the potential of an increasing interpretabilty of the mode. \cite{thekumparampil_attention-based_2018} propose a variation of this model, \textit{Attention-based Graph Neural Networks} (AGNN), where the \textit{relevance} is defined based on the cosine similarity.
		
		\paragraph{GraphUNet} \cite{gao_graph_2019} proposed GraphUNet, based on a new definition of Pooling layers. Convolutional layers in computer vision are normally used with \textit{Pooling layers}. This is because the convolutional filters are trained to detect the presence of a specific feature in a portion of the image, and if the feature is found, the result will have high positive values. The max pooling layer, makes a downsizing of the input and captures the highest values. By doing this, it generates a clear indication whether or not that particular feature was present.
		However, as it happens with traditional convolutions, the pooling layers are defined based on the regular pattern of the image. Defining a pooling layer on the graph domain could be useful for building better representations of hierarchical patterns.
		\cite{gao_graph_2019} proposed a new definition of pooling \textit{gPool}. The gPool layer makes a linear projection of the nodes features, and a $k$-max pooling selection. With the identifier of the selected nodes, it builds the new (reduced) adjacency matrix and feature matrix. \cite{gao_graph_2019} also proposes an unpooling layer, which rebuilds the original network, and used together they can build an encoder-decoder architecture. The benefit of such an architecture is that it lets the node embedding be built based on the hierarchical properties of their neighbourhood. 
		
		\paragraph{Autoencoders}
		
		The Autoencoder is an architecture in deep learning where the network tries to learn the input, but goes through a compressed state. The network can be divided into two elements:
		
		\begin{itemize}
			\item the encoder, which can be a regular stack of layers, that ends with a vectorised representation of the input,
			\item the decoder, where the representation received from the encoder will be sized up to the original form. 
		\end{itemize}
		
		The idea is that, if the network is able to reconstruct the original input with a small error margin, then most of the information from the original input is correctly compressed in the low dimension at the end of the encoder.
		
		\cite{kipf_variational_2016} proposed the Graph Autoencoder (GAE) and the Variational GAE (a probabilistic implementation of the GAE), where the encoder can be any way of building a node embedding, like a stack of GCN layers, and the decoder is (for the GAE) the reconstructed adjacency matrix, $\hat{A}$:
		
		\begin{equation*}\label{GAE}
		\hat{A} = \sigma(ZZ^T), \qquad \text{with} \quad Z = GCN(X,A)
		\end{equation*}
		
		Where $Z$ represents the embedding matrix built using the GCN and $\sigma$ is a logistic sigmoid function. The model use the inner product of the node embeddings to reconstruct a probabilistic adjacency matrix. This transforms the embedding into edge probabilities for the given node pairs. This is a particularly useful architecture for the task of \textit{link prediction}.
		
		In this paper, we train our models for this task in a transductive setting. We randomly remove some citation links for test and validation set, and evaluate how well the reconstructed adjacency matrix can predict those removed links. We use as encoders the GCN, GraphSAGE, GIN, GAT, AGNN, and GraphUNet layers defined above, which constitute the current state of the art in the field.

	\pagebreak
	\section{Topic Modelling}
	\label{sec:appendix_lda}
		\setcounter{table}{0}
		\setcounter{figure}{0}    
    \begin{table}[htbp]
    \caption{Ten most relevant words for each topic, with $\lambda$=0.5}
    \label{table:lda_top}
    \begin{tabular}{clllll}
    \textbf{Topic} & \textbf{Word 1} & \textbf{Word 2} & \textbf{Word 3} & \textbf{Word 4} & \textbf{Word 5} \\
    \textbf{1} & Technology & Innovation & Policy & Development & Economic \\
    \textbf{2} & Patents & Firm & Technological & Property & Innovative \\
    \textbf{3} & Theory & Epistemic & Philosophy & Explanation & Argue \\
    \textbf{4} & Academic & Universities & Scientists & China & India \\
    \textbf{5} & Education & Causal & Student & Open & Teaching \\
    \textbf{6} & Mathematics & Mathematical & Probability & Properties & Function \\
    \textbf{7} & Information & Data & Evaluation & Method & Measures \\
    \textbf{8} & Public & Decision & Making & Risk & Political \\
    \textbf{9} & Index & Performance & Bibliometric & Publications & Output \\
    \textbf{10} & Patents & Firms & Technological & Innovation & Patents \\
    \textbf{11} & Collaboration & International & Countries & Global & National \\
    \textbf{12} & Network & Networks & Analysis & Structure & Semantic \\
    \textbf{13} & Citation & Journals & Journal & Citations & Impact \\
    \textbf{14} & Media & Springer & Epistemology & Communication & Medical \\
    \textbf{15} & Century & Philosophical & Quantum & Natural & Objects \\
    \textbf{16} & Knowledge & Belief & Space & Organizational & Identity \\
    \textbf{17} & Mind & Biology & Causation & Evolution & History \\
    \textbf{18} & Realism & Law & The & Justification & Power \\
    \textbf{19} & Logic & Reasoning & Rationality & Cognition & Mental \\
    \textbf{20} & European & Social & Paradox & Politics & Engagement
    \end{tabular}
    \end{table}	
    
	\begin{figure}[htbp]
		\centering
		\includegraphics[width=\linewidth]{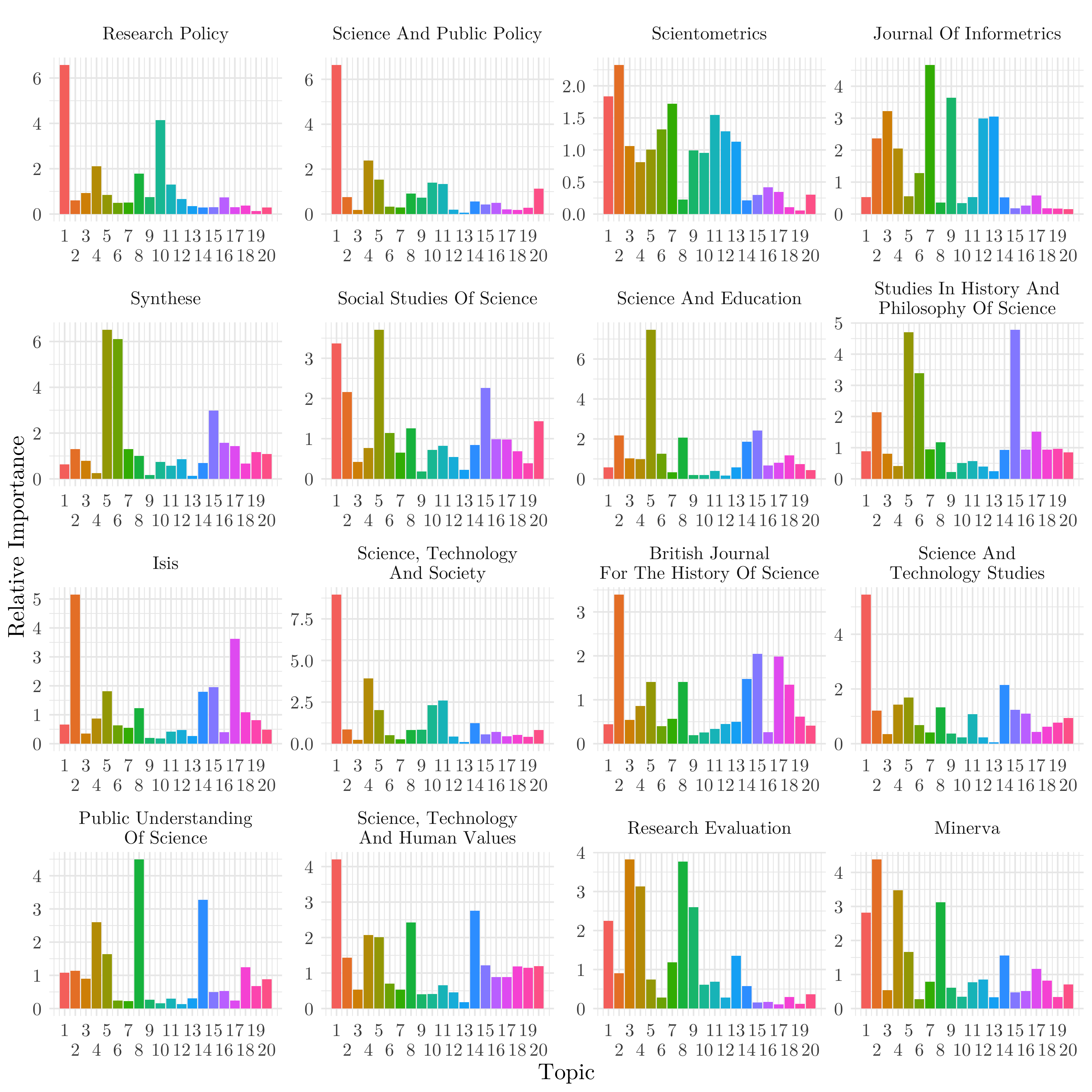}
		\caption{Relative importance of topics per journal.}
		\label{fig:topic_modeling_journals}
	\end{figure}
    
    \pagebreak
	\section{Frobenius Norm}\label{sec:frobenius_norm}
		\setcounter{table}{0}
		\setcounter{figure}{0}    
		
\begin{figure}[htbp]
			\centering
			\includegraphics[width=\linewidth]{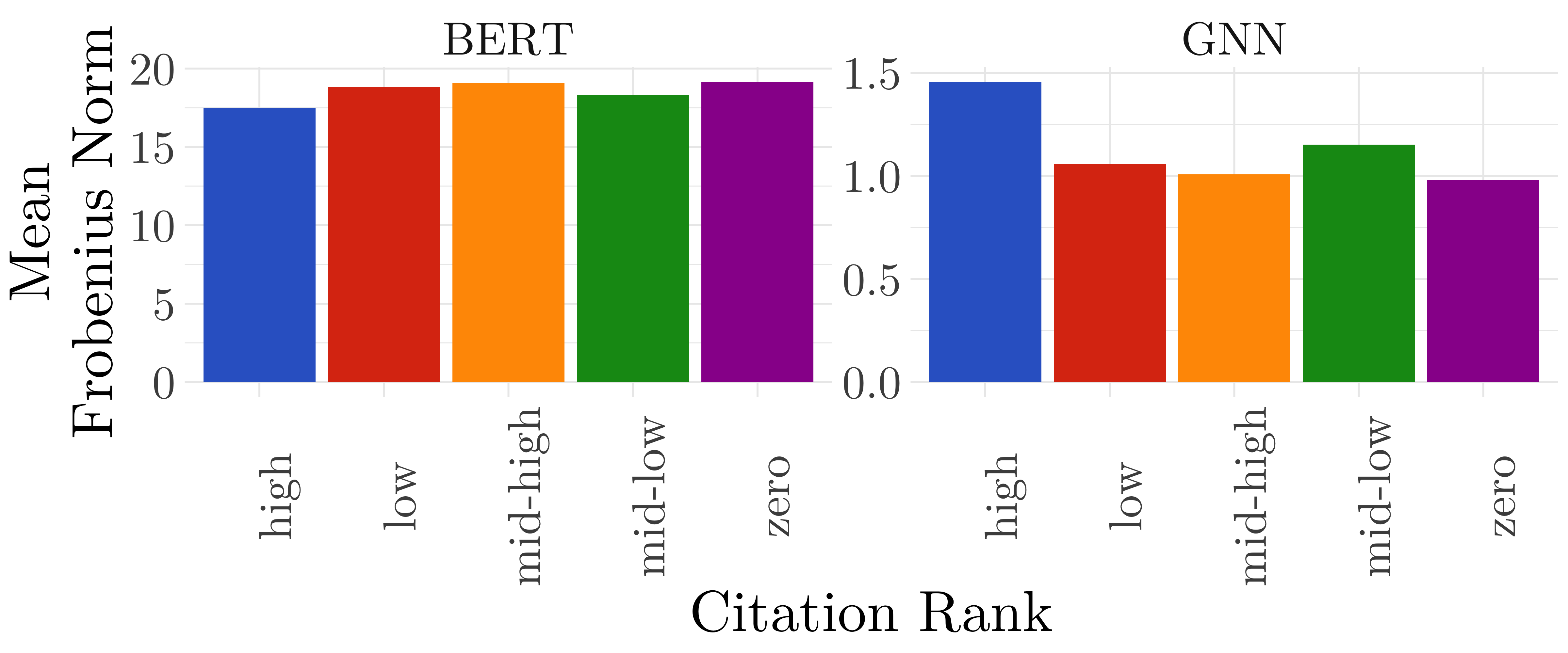}
			\caption{Frobenius Norm of articles by citation level.}
			\label{fig:frobeniousnorm}
		\end{figure}

		 In Figure \ref{fig:frobeniousnorm}, we can see the Frobenius norm of the article's embedding by their citation level. Those articles with lower citations than the 25\% threshold correspond to the group \textit{lower}, those between 25\% and 50\% to the \textit{mid-low}, between 50\% and 75\% to \textit{mid-high} and those articles with more than the 75\%  threshold belong to the group \textit{high}. For the GNN embedding we can see the higher norm of highly cited articles, while in the BERT embedding this does not sustain \footnote{Doc2Vec and LDA embeddings do not show significant differences between articles norms.}.
		
	\end{appendices}
	
\end{document}